\documentclass[twocolumn]{aa} 
\usepackage{graphicx}
\usepackage{longtable}        
\usepackage{natbib}             
\bibpunct{(}{)}{;}{a}{}{,}      
\usepackage{txfonts}
\begin{document}
   \title{On the nature of sn stars}

   \subtitle{I. A detailed abundance study}

   \author{C. Saffe\inst{1} \and
          H. Levato\inst{1}  }

   \institute{Instituto de Ciencias Astron\'omicas, de la Tierra y del Espacio (ICATE), C.C 467,
5400, San Juan, Argentina. Members of the Carrera del Investigador Cient\'{\i}fico, CONICET,
Consejo Nacional de Investigaciones Cient\'{i}ficas y T\'{e}cnicas de la Rep\'{u}blica Argentina
              \email{csaffe,hlevato@icate-conicet.gob.ar}
             }

   \date{Received xx/xx/xx; accepted xx/xx/xx}


  \abstract
   {The sn stars were first discoved by Abt \& Levato when studying the spectral types in different
open clusters. These stars present sharp Balmer lines, sharp metallic lines (C II, Si II, Ca II,
Ti II, Fe II), and broad coreless He I lines. Some of the sn stars seem to be related to CP stars.
Initially Abt \& Levato proposed a shell-like nature to explain the sn stars, although this scenario
was subsequently questioned. There is no general agreement about their origin.
We aim to derive abundances for a sample of 9 stars, including sn and non-sn stars,
  to determine the possible relation between sn and CP stars and compare their chemical
  abundances. That most sn stars belong to open clusters allows us to search for
  a possible relation with fundamental parameters, including the age and rotation.
  We also study the possible contribution of different effects to the broad He I lines observed
  in these stars, such as Stark broadening and the possible He-stratification. 
Effective temperature and gravity were estimated by Str\"omgren photometry and then refined by
    requiring ionization and excitation equilibrium of Fe lines.
    We derived the abundances by fitting the observed spectra with synthetic spectra using
   an iterative procedure with the SYNTHE and ATLAS9 codes.
We derived metallic abundances of 23 different chemical elements for 9 stars and
obtained low projected rotational velocities for the sn stars in our sample (vsini up to 69 km/s). 
We also compared 5 stars that belong to the same cluster (NGC 6475) and show that 
the sn characteristics appear in the 3 stars with the lower rotational velocity. 
However, the apparent preference of sn stars for objects with the lower vsini values 
should be taken with caution due to the small number of objects studied here.
We analysed the photospheric chemical composition of sn stars and show that approximately
$\sim$40$\%$ of them display chemical peculiarities (such as He-weak and HgMn stars) within a
range of temperature of 10300 - 14500 K. However, there are also sn stars with solar or nearly-solar
(i.e., non-CP) chemical composition.
We have studied the possible contribution of different processes to the broad He I lines present
in the sn stars. Although NLTE effects could not be completely ruled out, it seems that NLTE is not
directly related to the broad He I profiles observed in the sn stars.
The broad-line He I 4026 {\AA} is the clearest example of the sn characteristics in our sample.
We succesfully fit this line in 4 out of 7 sn stars by using the appropriate Stark
broadening tables, while small differences appear in the other 3 stars.
Studying the plots of abundance vs depth for the He I lines resulted in some sn stars 
probably being stratified in He. 
However, a further study of variability in the He I lines would help for
determining whether a possible non-uniform He superficial distribution could also play a role in
these sn stars.
We conclude that the broad He I lines that characterize the sn class could be modelled (at least
in some of these stars) by the usual radiative transfer process with Stark broadening,
without needing another broadening mechanism. The observed line broadening in sn stars
seems to be related to the "normal" He line formation that originates in these atmospheres.
}


   \maketitle
%

\section{Introduction}

The "sn" stars were characterized by \citet{abt-levato77} when classifying the
spectral types in the Orion OB1 association. These stars present
sharp Balmer line cores, sharp lines of metals (C II, Si II, Ca II, Ti II, Fe II),
and broad coreless lines of He I \citep[see also ][]{abt78}.
The authors introduced the "sn" designation because the spectra present
at the same time sharp (s) and nebulous (n) lines.
\citet{abt79} gave a list of 29 sn stars that belong to 12 open clusters and noted that
the sn characteristics could occur simultaneously with a variety of chemical peculiarities,
including He-weak, Bp(Si), and HgMn. 
These chemically peculiar (CP) stars present strong and/or weak intensities in the spectral
lines of certain chemical species.
The sn characteristics are detected in the spectra of B-type stars (B2-B9) with luminosity classes V-III
and in the HR diagram from the ZAMS to well above it. There have been no surveys of field stars
for sn spectra.

Abt \& Levato initially proposed that the effect is probably due to weak shells, generally related
to highly rotating stars. Shell stars show both strongly broadened photospheric lines and additional
narrow absorption lines \citep[see e.g. the reviews of][]{porter-rivinius03,rivinius06}.
In this context, the sharp lines are formed high in the atmosphere
(in the thin shell), while the diffuse or nebulous He I lines are formed lower and show the
rapid photospheric rotation. The observed rotational velocity of shell-like stars is usually
high, with vsini between 150-400 km/s \citep[see e.g. ][]{slettebak82,rivinius06}.
However, the CP stars are usually related to slow rotation.
\citet{charb-michaud88} determined that the maximum rotational velocities
allowing the gravitational settling of He are $\sim$75 km/s for HgMn stars and $\sim$100 km/s for FmAm.
Once the He abundance has decreased enough in the superficial
convection zones, the He convection zone disappears, and then, as the authors
explained, the abundance anomalies appear thanks to the radiative diffusion.
Observationally, the maximum rotational velocities measured for HgMn and Am stars
are 90 and 120 km/s, respectively \citep{wolff-preston78,abt-moyd73}.
More recent works also present similar maximum values for CP stars \citep[see e.g. ][]{abt-morrel95,abt09}.
\citet{mermilliod83} examined sn cluster stars and concluded that their main feature is also
a low projected rotational velocity: 12 out of 16 have vsini $<$ 50 km/s and all have vsini $<$ 100 km/s.
The coexistence of the sn characteristics with chemical peculiarities and the slow rotation,
favours a relationship with Bp stars.

Although most of the stars that present the sn characteristics
seem to present lower rotational velocities than shell stars,
a possible shell-like nature cannot be completely ruled out based solely on the rotational
properties. \citet{mermilliod83} show that the sn stars with the highest vsini in their sample
are HD 35502 and HD 36954 (285 and 190 km/s, respectively). These stars belong to the Orion association.
Also, \citet{neiner05} shows that the star HD 174512 is either a
shell Be star or a Herbig Ae/Be star, but with a rotational velocity of only 20 km/s.
Further studies are needed to confirm the real nature of this interesting object, which is a
candidate to be a shell star with a low rotational velocity.

There are several fundamental questions about the group of sn stars.
Previous works suggest a high frequency of CP stars among sn stars \citep[see e.g. Table 7 in][]{abt79},
which is also supported by their low projected rotational velocity \citep{mermilliod83}.
One of the aims of this work is to study the possibility that {\it{all}} sn stars were
in fact related to CP stars and search for common features in their abundance values.
A possible sn-CP relation would indicate that the sn characteristics are present in those
atmospheres with the physical conditions required by the diffusion to work efficiently.
Also we would like to determine T$_{\rm eff}$ and log g for the sn stars.
Some of the stars in our sample belong to the same cluster (NGC 6475), so the age and
the initial chemical composition are the same for all of them.
This allowed us to search for trends with different fundamental parameters.
We also included one sn star member of the cluster Melotte 20 and one from M45.
Some of these questions are addressed in this work, using a detailed abundance analysis
of a sample of sn (and non-sn) stars.

There is no general agreement about the possible contribution of different processes
to the broad He I lines observed in the sn stars.
For instance, in the temperature range of the sn stars, the Stark effect is an important
source of broadening \citep[e.g. ][]{shamey69,barnard74}.  In He-peculiar stars,
the atmosphere is possibly He-stratified, and this could affect the profiles of the He I lines
4471 {\AA} \citep{farthmann94} and 4026 {\AA} \citep{zboril05,catanzaro08}. Another aim of this work is
to determine the possible contribution of these effects (including NLTE) to the broad He I lines
observed in the sn stars. In other words, we study the possibility that the characteristic broadening
observed in the He I lines of sn stars could have originated, at least in part, in the atmospheres of
these stars.

In addition to the chemical pattern shown by the He-peculiar stars, there are a number of other
interesting phenomena present in the circumstellar regions of these stars.
\citet{shore-adelman81} reported variable UV lines in the spectrum of He-rich stars,
which are attributed to the magnetic control of the stellar mass outflows in these objects.
For the case of He-weak stars, \citet{brown84} started a systematic
survey of C IV line variations followed by other studies in the literature \citep{brown85,shore87}.
They studied a total of 15 He-weak stars, including both magnetic and non-magnetic stars, rapid and
slow rotators. Within this sample they detected enhanced absorption of C IV
only in three He-weak stars (HD 21699, HD 5737, and HD 79158) in which
the C IV resonance doublet and the magnetic fields are variable on the rotational time scale.
The authors interpret this result with magnetically controlled stellar winds among He-weak
stars \citep{shore87,shoreb87}.
Notably, these are the only three objects that share the sn designation in their sample
based on optical properties.
That these three stars display the C IV feature among many He-weak stars at
similar temperatures suggest some mechanism working in the sn stars but not in the 
other "normal" He-weak stars \citep{shore87}.

The initial observation of two additional He-weak sn stars (HD 21071 and HD 22136) shows no strong
C IV absorption in the spectra \citep{shore86,shore87}. However, it is possible that these stars
are C IV variable, similar to other He-weak stars, and a single observation corresponds
to their weak line phase \citep{shore86}.
More recently, \citet{shore04} have studied the presence of magnetically controlled circumstellar plasmas
between the hot members of the He-weak stars, including both sn and non-sn stars.
They used UV spectra from the International Ultraviolet Explorer satellite (IUE) which cover the C IV
and other important resonance lines. These lines provide a spectroscopic signature of the plasmasphere
through the variation in the C IV and Si IV resonance doublets.
For the first time, they discovered variable C IV in He-weak stars that do not belong to the sn subclass.
The UV line profile variations that are common between He-strong stars \citep[e.g. ][]{shore-adelman81,barker82}
seem to be more widespread among the He-weak stars. In other words, although the UV line profile
variation is not rectricted to only the He-weak sn stars, these works show the important result that the
He-weak sn stars could present magnetically controlled circumstellar plasmas detected in their UV spectra.

We describe the observational material in Section 2.
In Section 3 we describe the derivation of the temperature and gravity of the sn stars,
which are required to obtain an initial model atmosphere for our objects.
In Section 4 we explain the iterative method applied to
derive the abundances by fitting the observed spectra with synthetic ones.
We also discuss the possible NLTE effects present in our calculation.
In Section 5 we compare the abundance values of sn and CP
stars and discuss the possible contribution of different
physical processes to the broadening of He I lines.
Finally, we conclude the work with some remarks in Section 6.


\section{Observational material}

The stellar spectra of the sn stars were obtained from the ESO Science Archive Facility\footnote{Based
on data obtained from the ESO Science Archive Facility, http://archive.eso.org}
for the HARPS and UVES spectrographs, and from the Observatoire de Haute-Provence archive for 
ELODIE spectrograph\footnote{http://atlas.obs-hp.fr/elodie/}.
HARPS is a fibre-fed high-resolution echelle spectrograph \citep{mayor03}, installed at the 3.6m ESO
telescope at La Silla observatory. HARPS resolving power is $\sim$110000, and it uses two CCD EEV type 44-82
(4kx4k) detectors with 15 $\mu$m pixels. The spectral range covered is $\sim$3800-6800 A.
UVES is a two-arm cross-dispersed echelle spectrograph \citep{dekker00} attached to the VLT Kueyen UT2
telescope with the Nasmyth B focus. UVES resolving power is $\sim$ 80000 and the spectral range covers
3000-5000 A (blue) and 4200-11000 A (red). The CCD detector in the blue is an EEV (2kx4k), while in the red
there is an EEV chip mosaic (2kx4k) and an MIT/LL chip that has a higher efficiency in the
near-infrared. ELODIE is a cross-dispersed high-resolution echelle spectrograph attached to the
Observatoire de Haute-Provence 1.93m telescope. The spectral range covers 3850-6800 A with a resolving
power of 42000, split in 67 spectral orders. This spectrograph uses a 1024x1024 CCD SITe detector, with
24 $\mu$m pixels. The S/N of the spectra varies between 180 - 370, with an S/N average of $\sim$245
approximately.

The nine stars in our sample have V magnitudes in the range 4.3-6.2 mag and early-B spectral types. 
In Table \ref{sample.table} we present some observational data for each star.
The spectral types were taken from the Hipparcos database, while the sn (or non-sn) classification
of the stars were taken from the literature (see columns 5 and 6).
We included 6 sn stars in our sample \citep{abt79,mermilliod83,shore87} and the sn "candidate" HD 74146 
\citep{mermilliod83}. For comparison purposes, we also added two non-sn stars (HD 162630 and HD 162817) that belong to NGC 6475.
The reference for the sn classification is presented in the last column.

\begin{table}
\caption{Sample of stars studied in this work with spectral types 
taken from the Hipparcos database.}
\hskip -0.35in
\begin{tabular}{lccccc}
\hline
\hline
Star       & Cluster        & Instrument & Spectral & sn & Ref.\\
HD         &                &            & type \\
\hline
5737    &                & HARPS  & B7p, He-wk& Y & R3 \\ 
21071   & Melotte 20 675 & ELODIE & B7V& Y & R1,R2 \\
23950   & M45 3325       & ELODIE & B8, HgMn& Y & R1,R2 \\
74146   & IC 2391 16     & HARPS  & B4IV &(Y) & R2 \\
162586  & NGC 6475 56    & HARPS  & B8V& Y & R1,R2 \\
162630  & NGC 6475 63    & UVES   & B9III&N & R1\\
162678  & NGC 6475 141   & HARPS  & B9V &Y & R1,R2 \\
162679  & NGC 6475 77    & HARPS  & B9V &Y & R1,R2 \\
162817  & NGC 6475 110   & UVES   & B9.5V &N & R1\\
\hline
\end{tabular}
\tablebib{R1 \citep{abt79}, R2 \citep{mermilliod83}, R3 \citep{shore87}}
\label{sample.table}
\end{table}

By inspecting the spectra of our sample stars, the broad line He I 4026 {\AA}
is always present and is probably the clearest example of the sn characteristic.
We return to this topic in section 5.3.

The spectral lines of the stars were identified using a similar procedure
to previous works \citep[e.g. ][]{saffe2,saffe11}. 
The atomic line list and log gf data used in this work was basically taken from Fiorella Castelli's
linelist\footnote{http://wwwuser.oat.ts.astro.it/castelli/linelists.html},
with some changes and additions as described in \citet{castelli-hubrig04}. This linelist
includes the \citet{fuhr-wiese06} log gf data for several Fe I and Fe II lines,
usually referred to as FW06.
The line data and log gf values were updated according to 
\citet{adelman06} and \citet{zavala07}, in agreement with our previous abundance works
\citep[e.g. ][]{saffe11}.


\section{Atmospheric parameters}

We used the Str\"omgren uvby$\beta$ mean colours from \citet{hauck-mermilliod98},
together with the calibration of \citet{napi93}, to derive a first estimation of T$_{\rm eff}$
and log g. However, the photometric standard calibrations should be used with caution
in the case of CP stars (some of which are included in our sample), owing to their abnormal
colours and non-solar chemistry. \citet{adelman-rayle00} and \citet{netopil08} derived T$_{\rm eff}$
corrections for the Str\"omgren and other photometric systems, depending on the chemical
peculiarity. For the case of HgMn stars, the T$_{\rm eff}$ directly derived by Str\"omgren
photometry is overestimated, so that a correction is neccessary \citep{adelman-rayle00,netopil08}.
We corrected the derived values of the HgMn star HD 23950 according to
\citet{adelman-rayle00}, who found a systematic difference between parameters derived
from photometric and spectrophotometric techniques. They suggest a general correction
for HgMn stars that closely agrees with those derived by \citet{netopil08}.
For Am stars, \citet{netopil08} conclude that the known calibrations
for normal stars can be safely used with high accuracy ($\sim$ 150 K). 
\citet{paunzen02} show that the standard Str\"omgren calibration for normal
stars is also valid for $\lambda$ Boo stars.

In the case of He-weak stars, the T$_{\rm eff}$ derived by Str\"omgren is also overestimated
\citep[e.g.][]{netopil08}. \citet{leone-manfre97} determined the temperature and gravity accurately
for the He-weak star HD 5737 by using an iterative procedure.
They determined the fundamental parameters by fitting the H$\beta$ line using the program SYNTHE,
then they recalculated the He and metal abundances, derived a new ATLAS9 model
atmosphere, and restarted the process. In other words, they derived T$_{\rm eff}$
and log g but used non-solar abundances. For this object we adopt the
accurate parameters obtained by these authors (13600 K, 3.20 dex).
The T$_{\rm eff}$ derived using the Str\"omgren photometry for Ap stars should also be
corrected only if T$_{\rm eff}$ $>$ 9000 K, while no correction is needed for cooler Ap stars
\citep{netopil08}. Our sample includes the star HD 162817, which shows mild-Ap rather
than classical Ap characteristics (see next sections). The T$_{\rm eff}$ corrections are
not clearly adequate for this object so we prefer to use the original Str\"omgren values.

\begin{figure}
\centering
\includegraphics[width=6cm]{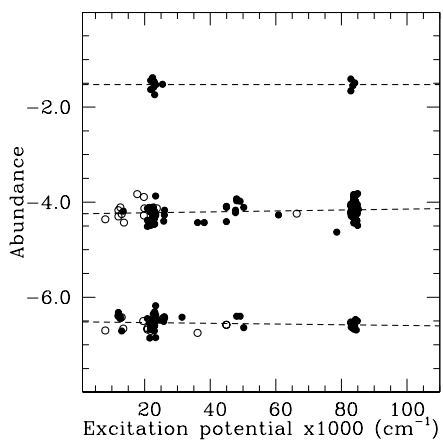}
\caption{Example of abundance vs excitation potential of Fe lines for the stars
HD 23950, HD 5737, and HD 162678 (from top to bottom). Empty and filled circles
correspond to Fe I and Fe II values, respectively. The straight lines show the 
position of a linear fit to the data. In this plot the abundance values have been
displaced vertically to avoid superposition.}
\label{fig.ep}
\end{figure}

We refined the T$_{\rm eff}$ to
achieve the excitation equilibrium condition, i.e. the independence between the
abundance values and excitation potential of the lines. In Figure \ref{fig.ep} we
present an example of abundance vs excitation potential of the Fe lines for the stars
HD 23950, HD 5737, and HD 162678. The figure also shows a linear fit to the abundance data.
Next we adjusted the surface gravity of the sn stars to get ionization equilibrium,
i.e. [Fe I/H]$=$[Fe II/H] within the observational errors. A similar strategy was applied
in previous works \citep[e.g. ][]{adelman-yuce10,saffe11,saffe-fundpar}. Finally, the adopted
values are shown in Table \ref{tefflogg}, together with the corresponding age of the cluster.

\begin{table}
\center
\caption{Fundamental parameters adopted for the stars in our sample.}
\vskip 0.1in
\begin{tabular}{lccccc}
\hline
\hline
 Star     &T$_{eff}$& log g & v sin i & Age & Ref. age\\
          & [K]     & [dex] & [km/s] & [Myr] & \\
\hline
HD 5737   & 13600 & 3.20 & 17 \\
HD 21071  & 14768 & 4.30 & 58 & 90  & R1 \\
HD 23950  & 12892 & 4.02 & 69 & 100 & R2 \\
HD 74146  & 14180 & 4.23 & 36 & 53  & R3 \\ 
HD 162586 & 12503 & 4.11 & 27 & 224 & R2 \\
HD 162679 & 10519 & 3.74 & 40 & 224 & R2 \\
HD 162630 & 10625 & 3.72 & 49 & 224 & R2 \\
HD 162678 & 10300 & 3.45 & 40 & 224 & R2 \\
HD 162817 & 10428 & 3.25 & 75 & 224 & R2 \\
\hline
\end{tabular}
\tablebib{For the adopted ages: R1 \citep{stauffer99}, R2 \citep{meynet93},
R3 \citep{barrado99}}
\label{tefflogg}
\end{table}

To determine the rotational velocities of the sn stars we fitted a synthetic spectrum
to $\sim$20 Fe II lines. We used the program SYNTHE \citep{synthe}, together with
the command {\it{broaden}} to reproduce the instrumental broadening of the spectrographs,
adopting a resolving power R of 110000, 80000, and 42000 for the HARPS, UVES, and ELODIE
spectrographs, respectively. The value of R is approximate, and thus the derived rotational
velocities should be taken with caution. The final v sin i values were obtained using the
average of the different Fe lines. The average dispersion is $\sim$0.9 km/s.
The rotational velocities are presented in the fourth column of Table \ref{tefflogg}. 
We agree with the conclusion of \citet{mermilliod83} that the sn stars
present low projected rotational velocities.

\section{Abundance analyses}

We used an iterative procedure to determine the abundances of chemical elements
in the atmospheres of the sn stars.
With the adopted values of T$_{\rm eff}$ and log g, we started by
computing a model atmosphere using the ATLAS9 \citep{kurucz93} code. This initial model
has [M/H]$=$0.0 i.e. solar metallicity values taken from \citet{solar}.
To determine the abundances we fitted a synthetic spectra to the different lines
using the program SYNTHE\footnote{http://kurucz.harvard.edu/programs.html}.
With the new abundance values, we derived a new model atmosphere and started
the process again.

We compared the synthetic and observed spectra numerically using the $\chi^{2}$
function (the cuadratic sum of the differences between both spectra).
The abundance is modified in steps of 0.01 dex for each spectral line, until a
minimum of $\chi^{2}$ is obtained. 
We present the abundances for each sn
star studied in this work in Tables \ref{tab1.abund}-\ref{tab3.abund}. 
In the ATLAS9 model atmospheres, for each chemical species the abundance is directly given
as log(N/N$_{T}$), where N is the number of absorbing atoms per unit volume and N$_{T}$ the
number of particles of both H and He. In other words, it is only necessary to specify the
ratio N/N$_{T}$ to determine certain abundance value.
The final values presented in Tables \ref{tab1.abund}-\ref{tab3.abund} correspond
to the average and rms ($\sigma$) derived from different spectral lines.
To show the possible overabundances or underabundances of the sn stars,
we also present the average using the square bracket notation, which denotes abundances
relative to the Sun, i.e. [N/N$_{T}$]=log(N/N$_{T}$)$_{Star}$-log(N/N$_{T}$)$_{Sun}$.
Solar abundance values log(N/N$_{T}$)$_{Sun}$ have been taken from \citet{solar}.
We caution that the error of the abundances $\sigma$ presented in these tables
correspond to the dispersion of the abundances derived from different lines and
do not include the error in the abundances of the Sun, in order to properly compare
with previous abundance works \citep[e.g.][]{zavala07,saffe11}. 
No uncertainty is quoted when only one line is present. 

In Table \ref{lines.table} we show a sample
of the line by line abundances, in this case for the star HD 162586.
The columns present the wavelength and abundance log(N/N$_{T}$) of the line,
log gf with their corresponding reference, the minimum $\chi^{2}$
derived, and the residual intensity of the isolated synthetic line (before rotational and
instrumental broadening).
The next column presents a flag (Y or N) if the line is used or not to calculate
the average abundance of the element, and finally a comment for each particular line.
The chemical species are identified at the beginning of the line lists using the
code "z.i", where z is the atomic number and i the ionization stage. For instance,
14.01 corresponds to Si II. The same code, z.i, is used to identify elements in most
figures in this work.
There is an on-line version of the complete Table \ref{lines.table} including a plot
for each line\footnote{http://icate-conicet.gob.ar/saffe/sn/Html/Salida7.html}.

\begin{table*}
\center
\caption{Derived abundances for the stars in our sample.
For each species, we present the abundance log(N/N$_{T}$), the abundance relative
to the Sun, [N/N$_{T}$], and the standard deviation $\sigma$ with the number of lines used (n).
The last column present the solar abundances log(N/N$_{T}$)$_{Sun}$ taken from \citet{solar}.  }
\begin{tabular}{lcccccccccc}
\hline
\hline
          & & HD      &               & & HD       &             & & HD& & \\
Species   & & 5737    &               & & 21071    &             & & 23950 &         & Sun    \\
\hline
  & log(N/N$_{T}$)&[N/N$_{T}$]&$\sigma$ (n) & log(N/N$_{T}$)&[N/N$_{T}$]&$\sigma$ (n) & log(N/N$_{T}$)&[N/N$_{T}$]&$\sigma$ (n) & log(N/N$_{T}$)$_{Sun}$ \\
\hline
He I      & -1.75 & -0.68 & 0.16 (7)  & -1.18 & -0.11 & 0.06 (9)  & -2.04 & -0.97 & 0.02 (4)  &  -1.07\\
C II      & -3.48 & 0.04 & 0.20 (6)    & -3.91 & -0.39 &  (1)     & -3.96 & -0.44 &  (1)     &  -3.52\\
Mg II     & -4.6 & -0.14 & 0.10 (5)    & -4.59 & -0.13 & 0.06 (4)  & -4.99 & -0.53 & 0.10 (2)   &  -4.46\\
Al I      &      &       &            &       &       &           & -4.93 & 0.64 &  (1)      &  -5.57\\
Al II     & -6.95 & -1.38 &  (1)     & -5.79 & -0.22 &  (1)     &       &      &            &  -5.57\\
Si II     & -4.65 & -0.16 & 0.19 (12) & -4.72 & -0.23 & 0.05 (3)  & -4.83 & -0.34 & 0.07 (5)  &  -4.49\\
P II      & -6.57 & 0.02 & 0.25 (2)   &       &       &           & -5.4 & 1.19 & 0.32 (2)    &  -6.59\\
S II      & -5.1 & -0.39 & 0.15 (20)  & -5.12 & -0.41 & 0.12 (2)  &      &      &             &  -4.71\\
Cl II     & -5.21 & 1.33 & 0.23 (8)   &       &       &           &      &      &             &  -6.54\\
Ca II     & -5.17 & 0.51 &  (1)      &       &       &           & -5.4 & 0.28 &  (1)       &  -5.68\\
Sc II     & -7.71 & 1.16 & 0.15 (2)   &       &       &           &      &      &             &  -8.87\\
Ti II     & -5.78 & 1.24 & 0.19 (50)  &       &       &           & -6.21 & 0.81 & 0.22 (14)  &  -7.02\\
Cr II     & -5.68 & 0.69 & 0.13 (35)  &       &       &           & -5.85 & 0.52 & 0.23 (5)   &  -6.37\\
Mn I      &       &      &            &       &       &           & -3.98 & 2.67 & 0.17 (3)   &  -6.65\\
Mn II     &       &      &            &       &       &           & -4.49 & 2.16 & 0.22 (20)  &  -6.65\\
Fe I      & -4.13 & 0.41 & 0.15 (13)  & -4.74 & -0.2 &  (1)      &       &      &            &  -4.54\\
Fe II     & -4.2 & 0.34 & 0.13 (106)  & -4.74 & -0.2 & 0.15 (14)  & -4.4 & 0.14 & 0.26 (20)   &  -4.54\\
Ni II     & -6.73 & -0.94 & 0.13 (2)  &       &       &           &       &      &            &  -5.79\\
Ga II     &       &      &            &       &       &           & -5.53 & 3.63 & 0.17 (3)   &  -9.16\\
Sr II     & -7.03 & 2.04 & 0.16 (3)   &       &       &           &       &      &            &  -9.07\\
Nd II     & -8.98 & 1.56 & 0.18 (4)   &       &       &           &       &      &            & -10.54\\
Hg II     &       &      &            &       &       &           & -5.73 & 5.18 &  (1)      & -10.91\\
\hline
\end{tabular}
\label{tab1.abund}
\end{table*}

\begin{table*}
\center
\caption{Derived abundances for the stars in our sample. We use the same notation as in Table \ref{tab1.abund}.}
\begin{tabular}{lcccccccccc}
\hline
\hline
          & & HD      &               & & HD       &             & & HD& & \\
Species   & & 74146    &               & & 162586    &             & & 162630 &         & Sun    \\
\hline
  & log(N/N$_{T}$)&[N/N$_{T}$]&$\sigma$ (n) & log(N/N$_{T}$)&[N/N$_{T}$]&$\sigma$ (n) & log(N/N$_{T}$)&[N/N$_{T}$]&$\sigma$ (n) & log(N/N$_{T}$)$_{Sun}$ \\
\hline
He I      & -0.78 & 0.29 & 0.07 (9)   & -0.78 & 0.29 & 0.05 (8)   & -1.23 & -0.16 & 0.08 (5)  &  -1.07\\
C II      & -3.44 & 0.08 & 0.24 (5)   & -3.27 & 0.25 & 0.34 (5)   & -3.5 & 0.02 &  (2)       &  -3.52\\
N II      &       &      &              & -3.69 & 0.43 &  (1)    &      &      &             &  -4.12\\
O I       &       &      &            & -3.5 & -0.29 & 0.01 (3)   &      &      &             &  -3.21\\
Mg I      &       &      &            & -4.07 & 0.39 & 0.14 (4)   &      &      &             &  -4.46\\
Mg II     & -4.87 & -0.41 & 0.12 (4)  & -4.55 & -0.09 & 0.06 (6)  & -4.36 & 0.1 & 0.06 (2)    &  -4.46\\
Al I      &       &       &           &       &       &           & -5.56 & 0.01 &  (1)      &  -5.57\\
Al II     & -6.2 & -0.63 &  (1)      & -5.69 & -0.12 &  (1)     & -5.26 & 0.31 &  (1)      &  -5.57\\
Si II     & -4.92 & -0.43 & 0.17 (6)  & -4.59 & -0.09 & 0.05 (9)  & -4.42 & 0.07 & 0.08 (5)   &  -4.49\\
S II      & -4.68 & 0.03 & 0.18 (10)  & -4.66 & 0.05 & 0.07 (7)   &       &      &            &  -4.71\\
Ca II     & -6.45 & -0.77 &  (1)     & -5.59 & 0.09 &  (1)      & -5.66 & 0.02 &  (1)      &  -5.68\\
Sc II     &       &       &           & -9.02 & -0.15 &  (1)     &       &      &            &  -8.87\\
Ti II     &       &       &           & -6.84 & 0.18 & 0.21 (20)  & -6.98 & 0.04 & 0.18 (12)  &  -7.02\\
Cr II     &       &       &           & -6.25 & 0.12 & 0.11 (16)  & -6.23 & 0.14 & 0.24 (8)   &  -6.37\\
Fe I      &       &       &           & -4.41 & 0.13 & 0.12 (15)  & -4.44 & 0.1 & 0.14 (7)    &  -4.54\\
Fe II     & -5.22 & -0.68 & 0.22 (13) & -4.57 & -0.03 & 0.14 (60) & -4.41 & 0.13 & 0.11 (16)  &  -4.54\\
Ni II     &       &       &           & -5.79 & 0 & 0.15 (3)      & -5.41 & 0.38 & 0.29 (3)   &  -5.79\\
Sr II     &       &       &           & -8.66 & 0.41 & 0.15 (2)   & -8.32 & 0.75 & 0.14 (2)   &  -9.07\\
\hline
\end{tabular}
\label{tab2.abund}
\end{table*}

\begin{table*}
\center
\caption{Derived abundances for the stars in our sample. We use the same notation as in Table \ref{tab1.abund}.}
\begin{tabular}{lcccccccccc}
\hline
\hline
          & & HD      &               & & HD       &             & & HD& & \\
Species   & & 162678    &               & & 162679    &             & & 162817 &         & Sun    \\
\hline
  & log(N/N$_{T}$)&[N/N$_{T}$]&$\sigma$ (n) & log(N/N$_{T}$)&[N/N$_{T}$]&$\sigma$ (n) & log(N/N$_{T}$)&[N/N$_{T}$]&$\sigma$ (n) & log(N/N$_{T}$)$_{Sun}$ \\
\hline
He I      & -1.2 & -0.13 & 0.11 (8)   & -1.2 & -0.13 & 0.05 (7)   & -1.43 & -0.36 &  (1)     &  -1.07\\
C II      & -3.9 & -0.38 & 0.08 (4)   & -3.97 & -0.45 & 0.07 (3)  &       &       &           &  -3.52\\
O I       & -3.52 & -0.31 & 0.02 (2)  & -3.49 & -0.28 & 0.01 (3)  &       &       &           &  -3.21\\
Mg I      & -4.12 & 0.34 & 0.17 (3)   &       &       &           & -3.57 & 0.89 & 0.23 (2)   &  -4.46\\
Mg II     & -4.53 & -0.07 & 0.12 (5)  & -4.51 & -0.05 & 0.09 (5)  & -4.21 & 0.25 & 0.04 (2)   &  -4.46\\
Al I      & -5.85 & -0.28 &  (1)     & -5.74 & -0.17 &  (1)     &       &      &            &  -5.57\\
Al II     & -5.65 & -0.08 &  (1)     & -5.63 & -0.06 &  (1)     & -5.68 & -0.11 &  (1)     &  -5.57\\
Si II     & -4.54 & -0.05 & 0.06 (7)  & -4.56 & -0.07 & 0.08 (8)  & -4.49 & 0 & 0.09 (6)      &  -4.49\\
Ca II     & -5.66 & 0.02 &  (1)      & -5.6 & 0.08 &  (1)       & -5.13 & 0.55 &  (1)      &  -5.68\\
Sc II     & -9.16 & -0.29 & 0.06 (2)  & -9.32 & -0.45 & 0.20 (2)   & -8.68 & 0.19 &  (1)      &  -8.87\\
Ti II     & -6.87 & 0.15 & 0.23 (31)  & -6.74 & 0.28 & 0.23 (34)  & -6.35 & 0.67 & 0.23 (21)  &  -7.02\\
Cr II     & -6.28 & 0.09 & 0.16 (23)  & -6.15 & 0.22 & 0.23 (26)  & -6.13 & 0.24 & 0.20 (12)   &  -6.37\\
Fe I      & -4.49 & 0.05 & 0.11 (14)  & -4.39 & 0.15 & 0.16 (24)  & -4.05 & 0.49 & 0.25 (11)  &  -4.54\\
Fe II     & -4.56 & -0.02 & 0.14 (49) & -4.44 & 0.1 & 0.17 (53)   & -4.25 & 0.29 & 0.19 (19)  &  -4.54\\
Ni II     & -5.92 & -0.13 & 0.01 (2)     & -5.9 & -0.11 & 0.20 (2)    & -5.51 & 0.28 & 0.38 (2)   &  -5.79\\
Sr II     & -9.2 & -0.13 & 0.01 (2)   & -9.12 & -0.05 & 0.02 (2)  & -8.17 & 0.9 &  (1)       &  -9.07\\
\hline
\end{tabular}
\label{tab3.abund}
\end{table*}

\begin{table*}
\center
\caption{Sample of the line by line abundance table.}
\begin{tabular}{cccccccl}
\hline
\hline
Lambda & Abund & log gf & Ref & $\chi^{2}$ & Int & AbAccept & Comment\\
\hline
14.01 Si II \\
 3853.66 &  -4.50 & -1.440 & LA &  0.002441  &  0.3534 & Y \\
 3856.02 &  -4.57 & -0.490 & LA &  0.006904  &  0.2570 & Y \\
 3862.60 &  -4.60 & -0.740 & LA &  0.004232  &  0.2811 & Y \\
 4075.45 &  -4.63 & -1.400 & SG &  0.001527  &  0.8158 & Y \\
 4076.78 &  -4.59 & -1.670 & SG &  0.000888  &  0.8748 & Y \\
 4128.05 &  -4.58 &  0.380 & LA &  0.004738  &  0.3695 & Y \\
 4130.89 &  -4.60 &  0.530 & LA &  0.002375  &  0.3549 & Y \\
 4673.28 &  -4.48 & -0.710 & KX &  0.001259  &  0.9896 & Y & Weak and blended with Fe I 4673.2\\
 5041.02 &  -4.64 &  0.290 & SG &  0.001421  &  0.5269 & Y \\
\hline
 Si II Avg &  -4.59 & +/- 0.05 (9)\\
\hline
16.01 S II \\
 4153.07 &  -4.71 &  0.620 & WS &  0.001254  &  0.8554 & Y \\
 4162.66 &  -4.68 &  0.780 & WS &  0.001783  &  0.8287 & Y \\
 4415.34 &  -2.97 & -1.080 & KP &  0.006496  &  0.8903 & N & No fit\\
 4991.97 &  -4.60 & -0.650 & WS &  0.000846  &  0.9455 & Y \\
 5014.04 &  -4.68 &  0.030 & KX &  0.002286  &  0.9011 & Y \\
 5032.43 &  -4.61 &  0.180 & WS &  0.000620  &  0.8455 & Y \\
 5212.62 &  -4.77 &  0.240 & WS &  0.001127  &  0.9439 & Y \\
 5320.72 &  -4.56 &  0.460 & WS &  0.001090  &  0.9052 & Y \\
\hline
  S II Avg & -4.66 & +/- 0.07 (7)\\
\hline
20.01 Ca II \\
 3933.66 &  -5.59 &  0.130 & WM &  0.019829  &  0.1294 & Y & Intense line \\
\hline
  Ca II Avg & -5.59 &  (1)\\
\hline
\end{tabular}
\tablefoot{The chemical species are identified at the beginning of the line lists using the
code "z.i" where z is the atomic number and i the ionization stage. For instance,
14.01 corresponds to Si II. The same code z.i is used to identify elements in the figures
of this work.}
\label{lines.table}
\end{table*}

\begin{figure}
\centering
\includegraphics[width=8cm]{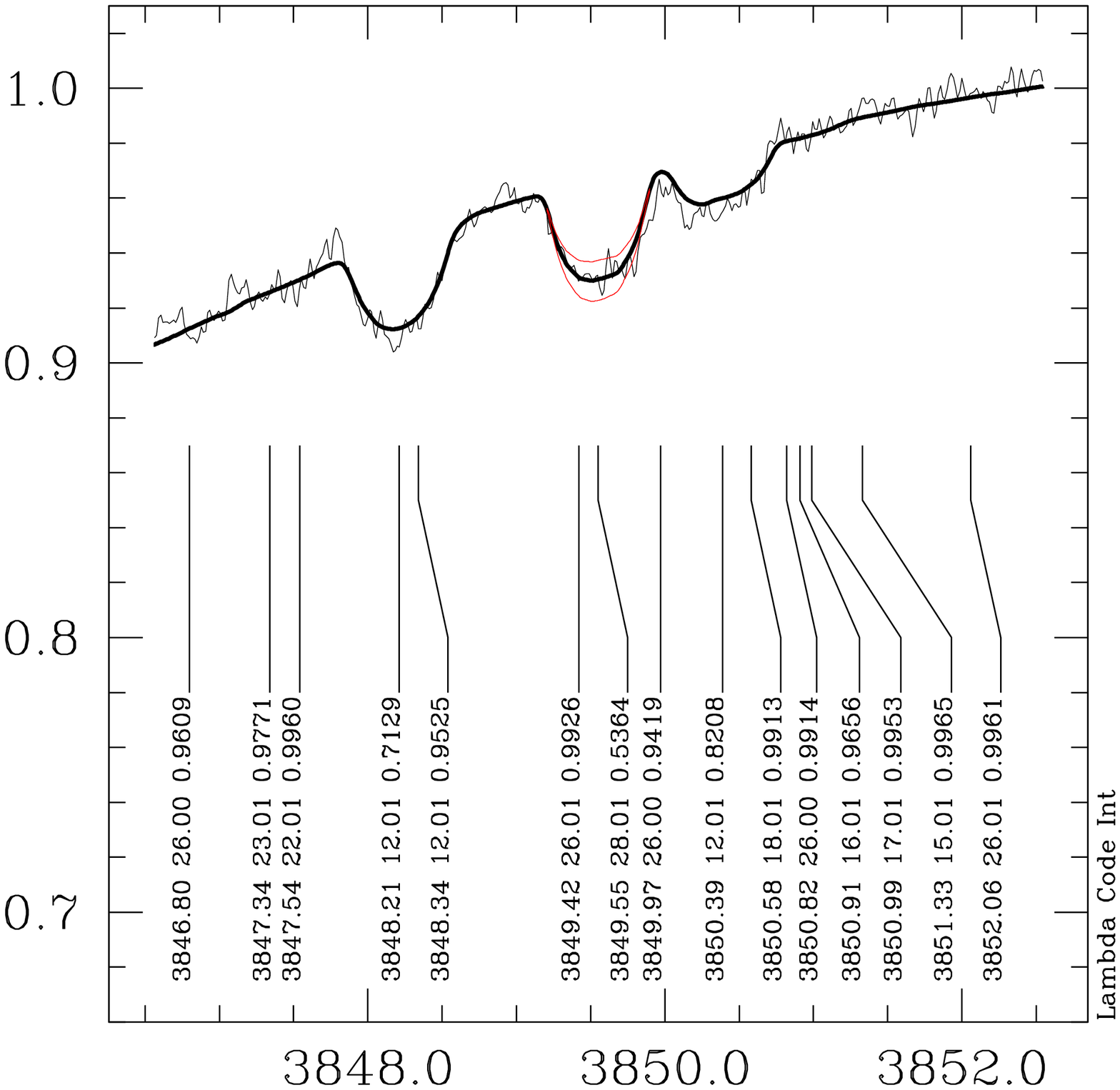}
\includegraphics[width=8cm]{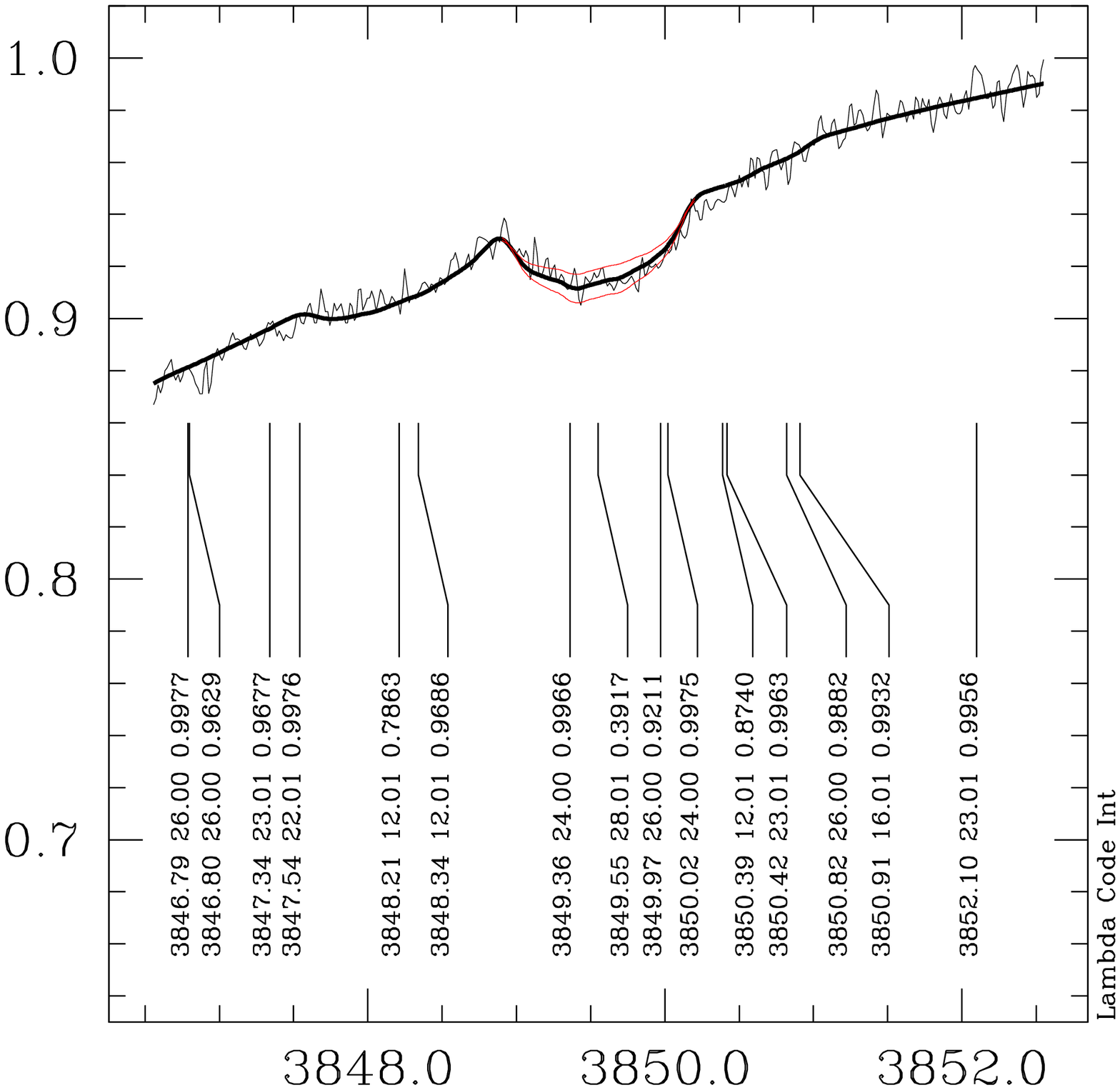}
\caption{Comparison of the spectral region near the Ni II 3849 line for
HD 162586 (upper panel) and HD 162630 (lower panel). The observed and synthetic
spectra are shown by thin and thick lines, respectively.
The 2 thin red curves correspond to $\pm$1$\sigma$ abundance variation 
of the synthethic spectra. }
\label{niii.fig}
\end{figure}

Figure \ref{niii.fig} shows two examples of synthetic fits of the same spectral region
near the line Ni II 3849.55 for the stars HD 162586 and HD 162630 (upper and lower panels,
respectively).
The synthethic lines are identified in the figure using the notation "Lambda Code Int", i.e.
the wavelength of the line in {\AA}, the code z.i of the element and the relative intensity of
the line. For instance, 3849.55 28.01 0.3917 correspond to the line 3849.55 {\AA} of Ni II,
with a relative intensity of 0.3917.
The two thin red curves correspond to $\pm$1$\sigma$ abundance variation in the
synthethic spectra, where $\sigma$ is the standard deviation derived in this
case from different Ni II lines. Both the $\pm$1$\sigma$ abundance curves and the notation
for identifying the synthethic lines are used in most figures of this work.
In the upper panel, the lines Mg II 3848 {\AA}, Ni II 3849 {\AA}, and Mg II 3950 {\AA}
are present (with some blends), while in the lower panel a higher vsini smooths
the region and strongly blends the lines. In both cases the synthetic spectra fit
well the observed spectra. 

\begin{figure}
\centering
\includegraphics[width=8cm]{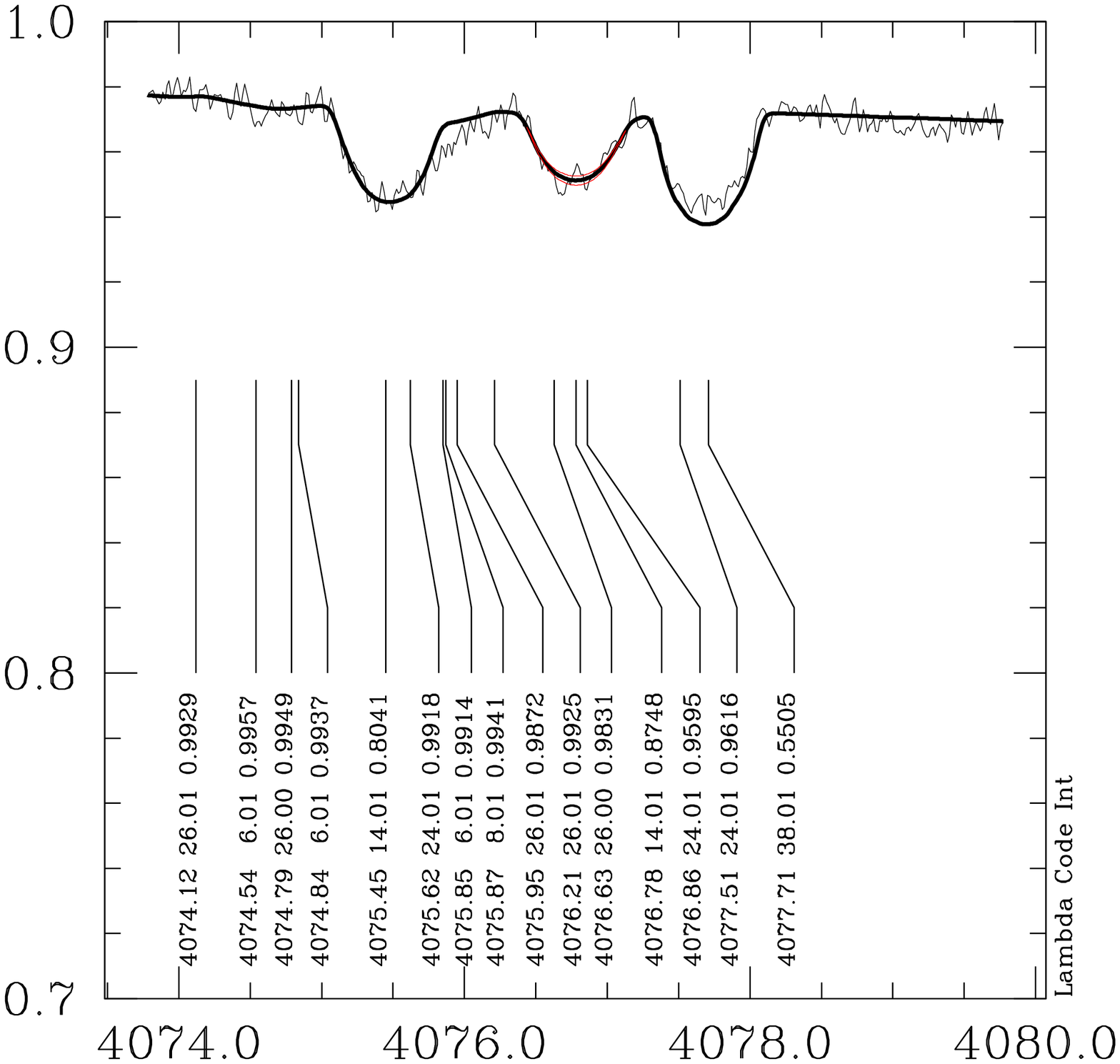}
\includegraphics[width=8cm]{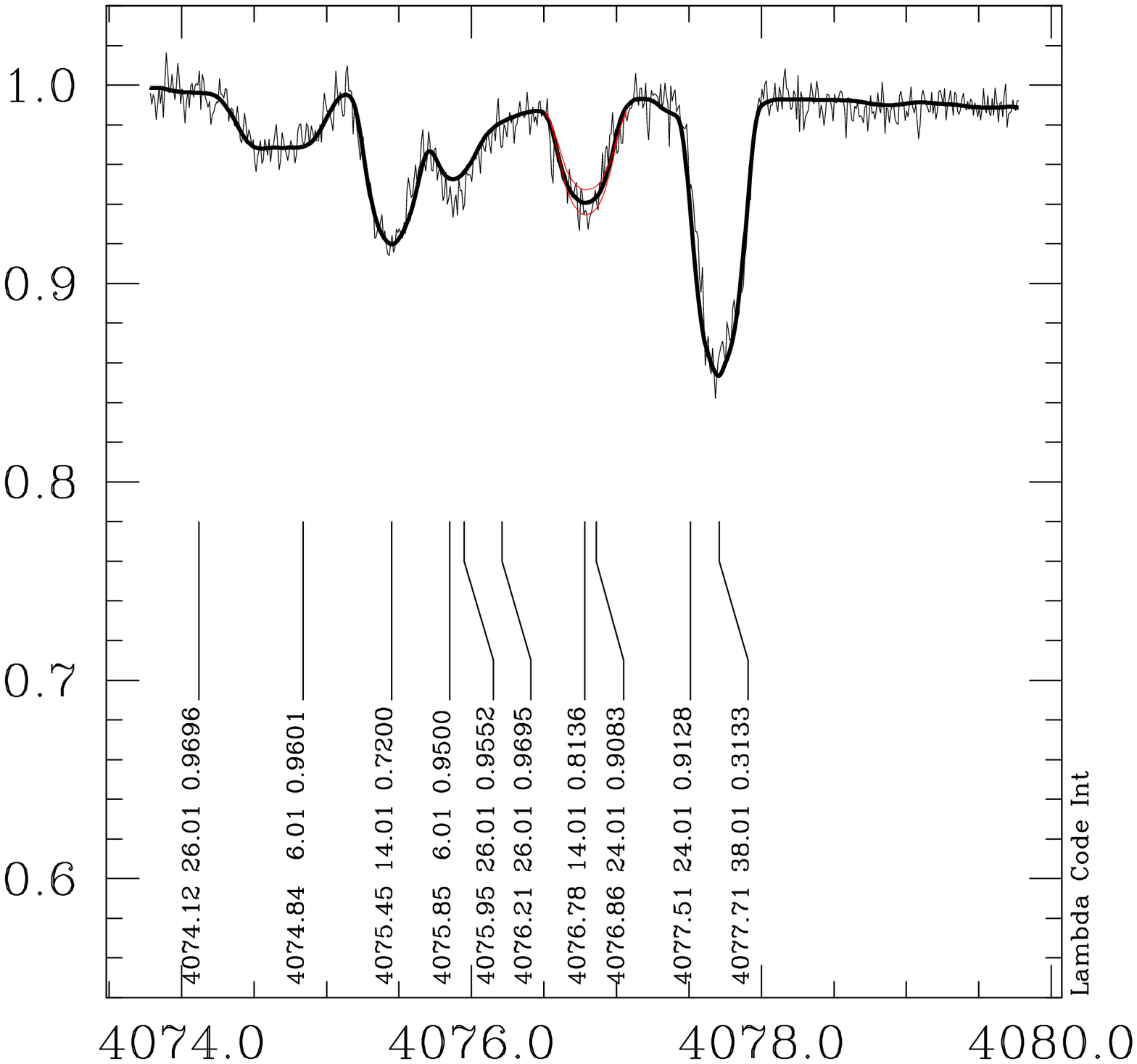}
\caption{Comparison of the spectral region near the line Si II 4076 for
HD 162586 (upper panel) and HD 5737 (lower panel). The observed and synthetic
spectra are shown by thin and thick lines, respectively.
The 2 thin red curves correspond to $\pm$1$\sigma$ abundance variation 
of the synthethic spectra. }
\label{siii.fig}
\end{figure}

In Figure \ref{siii.fig} we show two plots
corresponding to the spectral region near the line Si II 4076, for HD 162586 and
HD 5737 (upper and lower panels, respectively). In the upper panel there are three
intense lines (Si II 4075 {\AA}, 4076 {\AA}, and Ni II 4077 {\AA}) while in the lower panel there
are also C II lines (such as 4074.8 {\AA} and 4075.8 {\AA}) blended with other less
intense lines. We see that the use of synthetic spectra allows the modelling of
spectral lines even for rotating stars where some blends are present.

The uncertainty in the derived abundance values have different sources.
The effective temperature and gravity were first estimated by using the Str\"omgren photometry
and corrected for CP stars. This estimation presents average dispersions of $\sim$200 K and 
$\sim$0.2 dex for T$_{\rm eff}$ and log g, respectively \citep{adelman-rayle00,netopil08}.
We estimate similar dispersion values for T$_{\rm eff}$ and log g by achieving the
excitation and ionization equilibrium.
The total dispersion from the combination of photometry and equilibrium conditions
(assuming that they are independent) amount to $\sim$300 K in T$_{\rm eff}$ and $\sim$0.3 dex
in log g, approximately.
We estimate that the uncertainty in log gf values may cause average differences of about 0.09 dex
($\sim$4.7\%) in the calculated metallicity.
Finally, to provide an estimation of the "typical" uncertainty and give an idea of the sensitivity
of our results, raising the temperature of the sn sample 300 K ($\sim$3\%) increases the
average abundance by $\sim$0.07 dex (3.2\%), and raising the surface gravity by
0.3 dex ($\sim$21\%) increases the average abundance by $\sim$0.01 dex ($<$1\%).

\subsection{NLTE effects}

Most of the lines analysed here are weak lines formed deep in the atmosphere where
LTE should prevail. They are well-suited to abundance determinations. We have also
included data for hyperfine splitting for the selected transitions when relevant, using
the linelist {\it{gfhyperall.dat}}\footnote{http://kurucz.harvard.edu/LINELISTS/GFHYPERALL/}
However, the smearing out of spectra by stellar rotation clearly prevents us from detecting
signatures of hyperfine splitting and isotopic shifts.

Departures from LTE are more pronounced in stars with high temperature and with low gravity
and metallicity.  For instance, the reduction of surface gravity results in a decreased
efficiency of collisions with electrons and hydrogen atoms, reducing the thermalizing
effect, which leads to stronger NLTE effects \citep[see e.g. ][]{gratton99}.
It is not an obvious task to disentangle non-LTE effects from other sources of uncertainty
in abundance analyses. Subsequently, some particular transitions should be taken with caution.
In the line-by-line abundance table, we give a warning of "Possible NLTE effects" for those lines
where we suspect the presence of these effects. The case of possible NLTE effects, particularly
for the He I lines, is considered in the next sections (see sect. 6.1).

For instance, for C II both weak and intense lines show pronounced NLTE effects \citep{przybilla11}
except for C II 5145 {\AA} and the other members of the multiplet. It is possible to detect
NLTE effects by comparing the abundance values derived from C II 5145 (which is not affected by NLTE)
and abundances from intense C II lines (which present NLTE effects).
For O I, NLTE effects are expected, especially in the near-IR triplet O I 7771 {\AA} and the
other lines of the same multiplet \citep[see e.g.][]{sitnova13,przybilla11}.
\citet{przybilla01} evaluated the NLTE line formation of neutral and singly-ionized Mg,
and concluded that the NLTE effects for Mg II are usually small except for the 
intense line Mg II 4481 {\AA}. This line systematically yields higher abundance values than
the other Mg II lines, between 0.2-0.8 dex for stars with T$_{\rm eff}$ in the range
9100-9550 K. Most of the Si lines are also affected in stars with T$_{\rm eff}$ higher
than 15000 K \citep[e.g.][]{przybilla11}.
 

\section{Discussion}

\subsection{Abundances of sn stars vs. CP stars}

\begin{figure*}
\centering
\includegraphics[width=15.0cm]{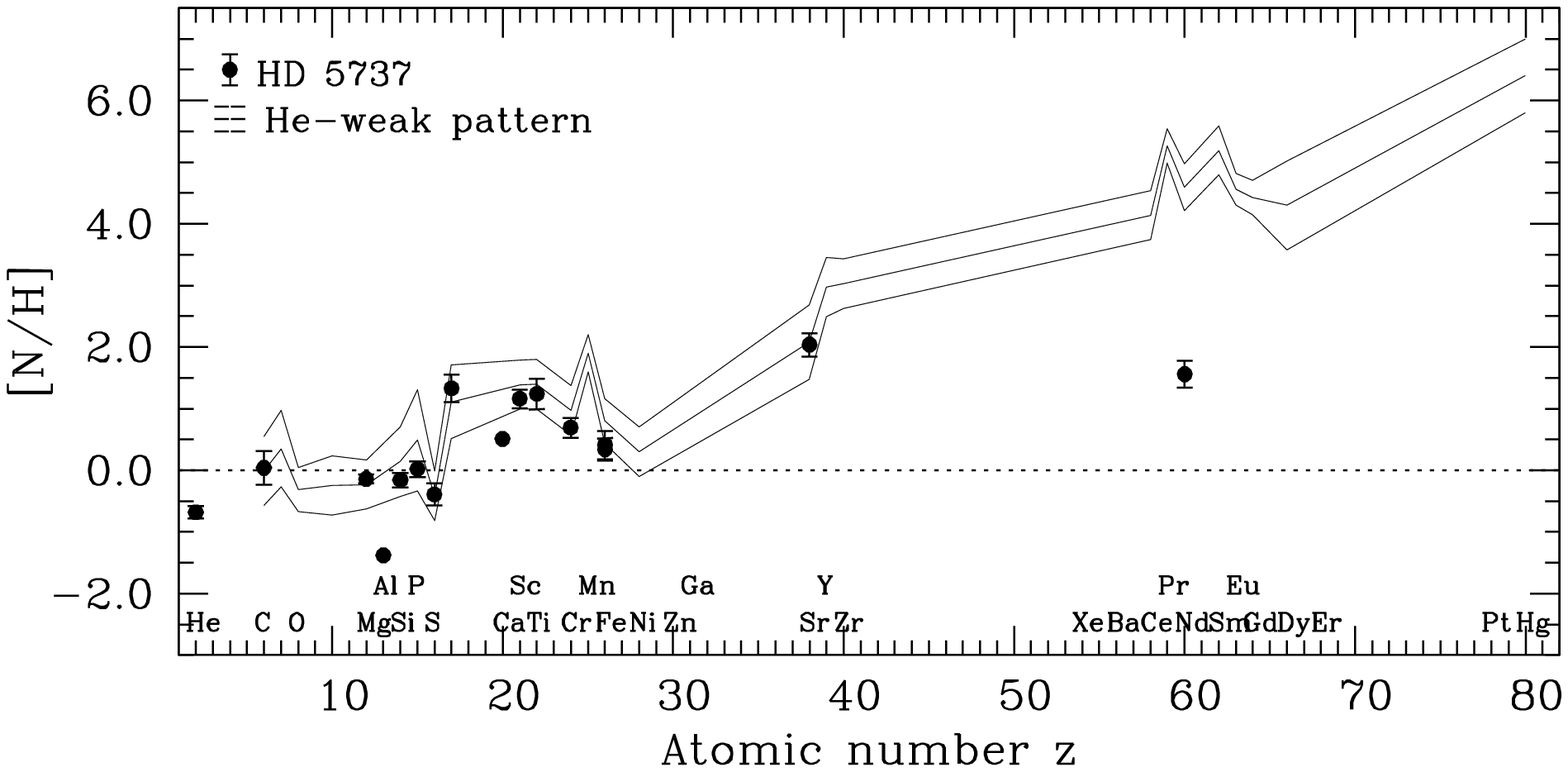}
\vskip -8cm
\includegraphics[width=15.0cm]{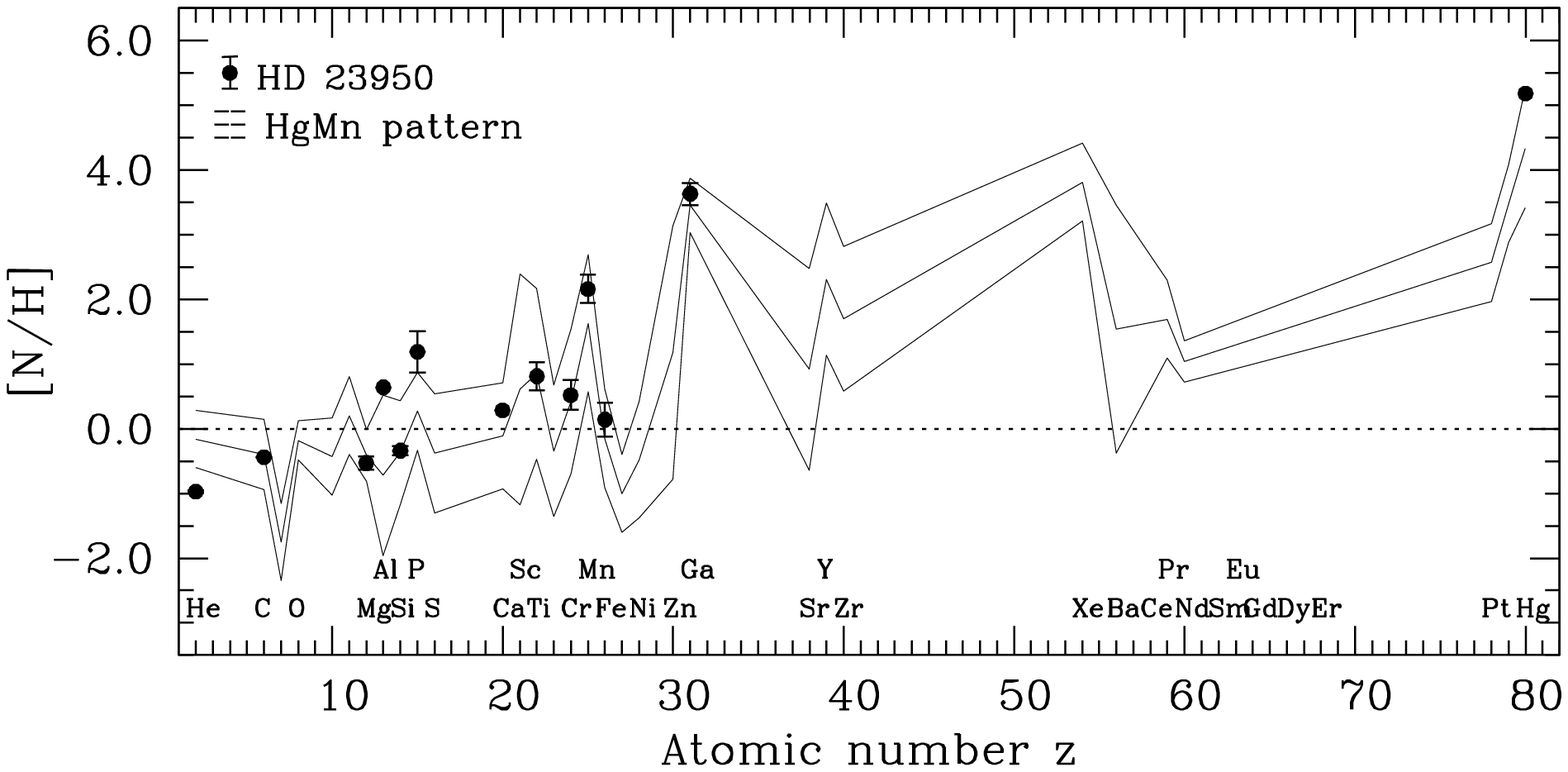}
\vskip -8cm
\includegraphics[width=15.0cm]{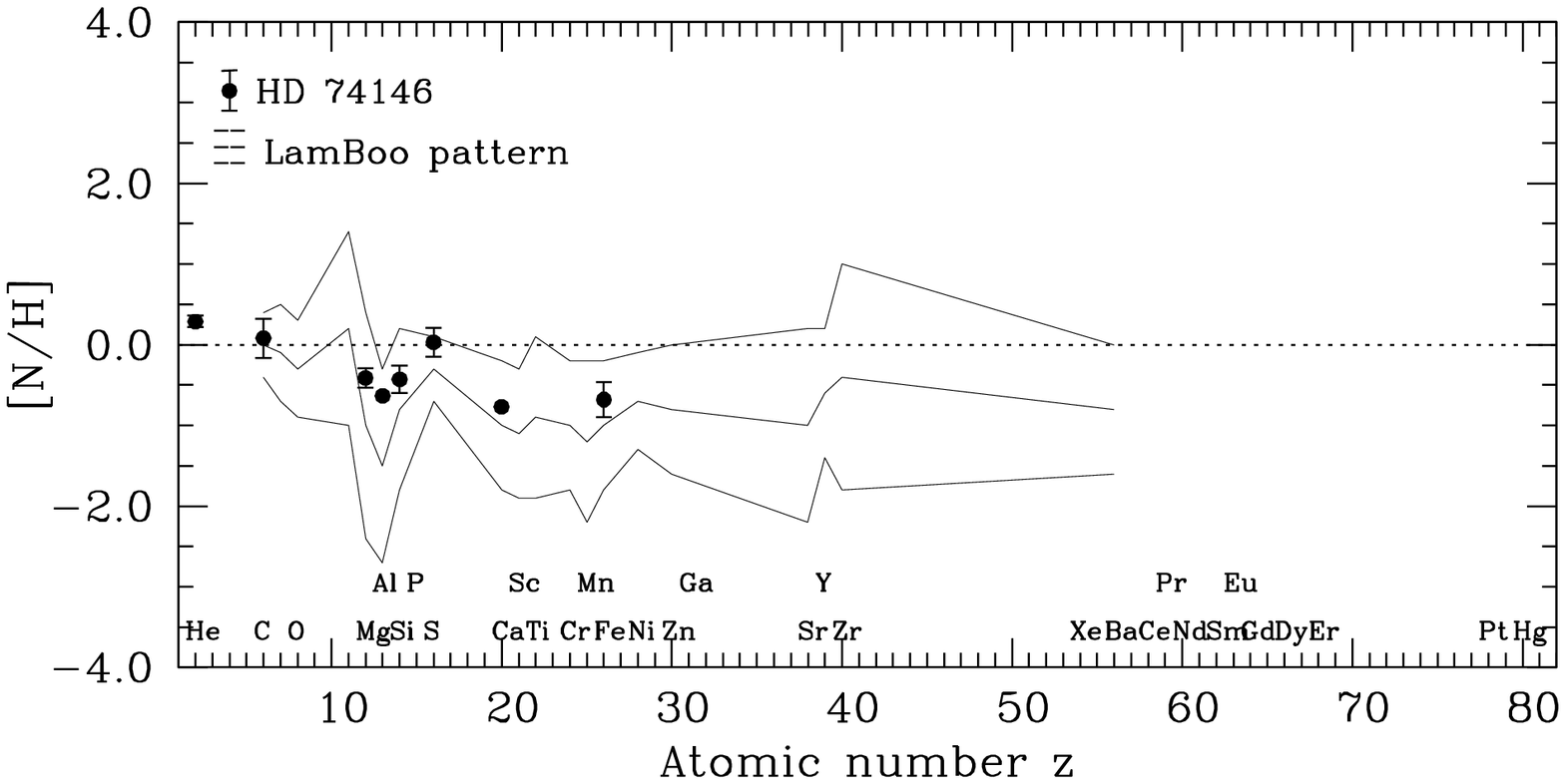}
\vskip -7cm
\caption{Comparison of abundance patterns as function of the atomic number.
Upper panel: HD 5737 (filled circles) vs. He-weak pattern (the 3 continuous lines).
Middle panel: HD 23950 (filled circles) vs. HgMn pattern (the 3 continuous lines).
Lower panel: HD 74146 (filled circles) vs. $\lambda$ Boo pattern (the 3 continuous lines).
See text for more details.}
\label{range.fig1}
\end{figure*}

\begin{figure*}
\centering
\includegraphics[width=15.0cm]{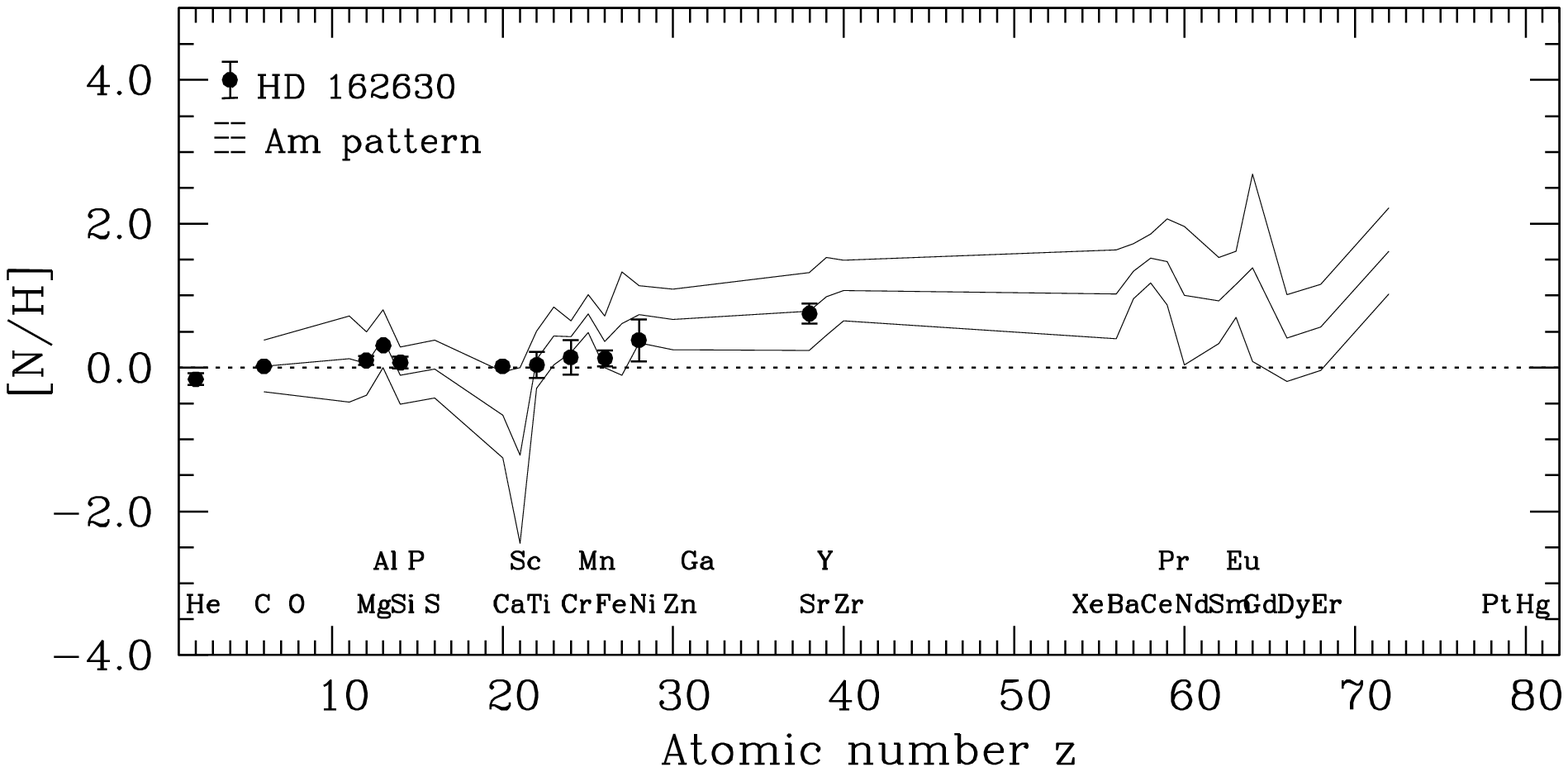}
\vskip -8cm
\includegraphics[width=15.0cm]{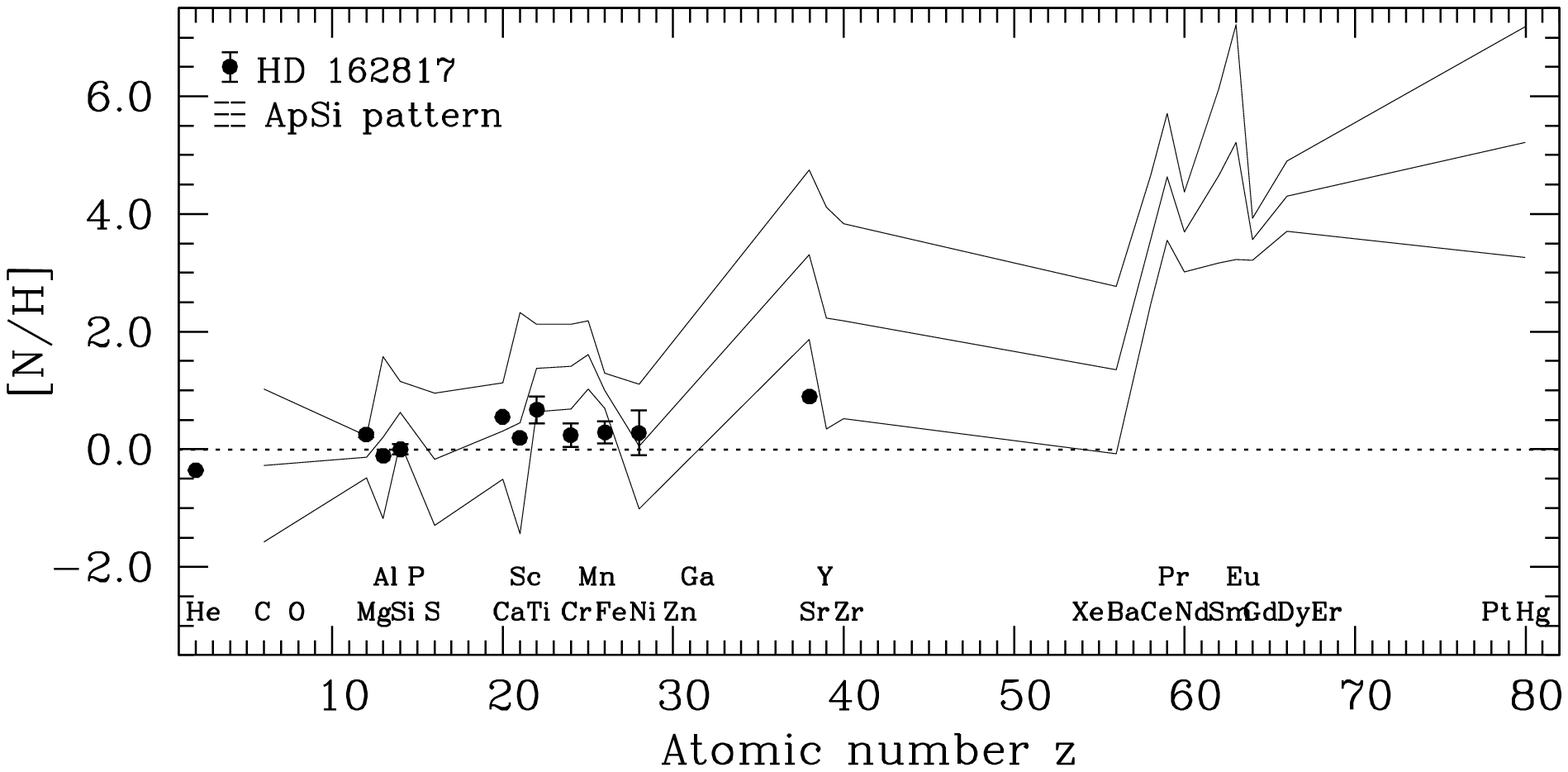}
\vskip -7cm
\caption{Comparison of abundance patterns as function of the atomic number.
Upper panel: HD 162630 (filled circles) vs. Am pattern (the 3 continuous lines).
Lower panel: HD 162817 (filled circles) vs. ApSi pattern (the 3 continuous lines).
See text for more details.}
\label{range.fig2}
\end{figure*}

We have determined the abundances of nine stars (including 6 sn stars, 1 sn candidate, and
2 non-sn stars) using synthetic spectra.
To compare these values with the chemical abundances of CP stars, we
present plots of abundance vs atomic number in Figures \ref{range.fig1} and \ref{range.fig2}. 
The plots present abundance values relative to the Sun, i.e. values higher 
than zero are overabundant, while lower than zero are deficient.
The CP stars present a range of possible abundance values rather than a single value for each
chemical species. Their abundance pattern is presented
as a "region" clearly demarcated by three continuous lines 
in Figures \ref{range.fig1} and \ref{range.fig2}.
For the HgMn pattern we used 15 HgMn stars listed in \citet{adelman06}, with the addition of
the HgMn star HD 165640 \citep{castelli-hubrig04}.
For the He-weak range we used four stars with abundances taken from the literature \citep{alonso03}, 
while for the $\lambda$ Boo pattern we used 12 stars taken from \citet{heiter02}.
For the ApSi pattern, we used five stars taken from literature \citep{lop01,lop02,lop03,alba02}.
On the other hand, Figures \ref{range.fig1}-\ref{range.fig2} show
the abundance values derived in this work for our sample stars and
the standard deviation of these values.

Next, some comments follow about individual stars, based mainly on the
abundance values derived.

\begin{figure}
\centering
\includegraphics[width=8cm]{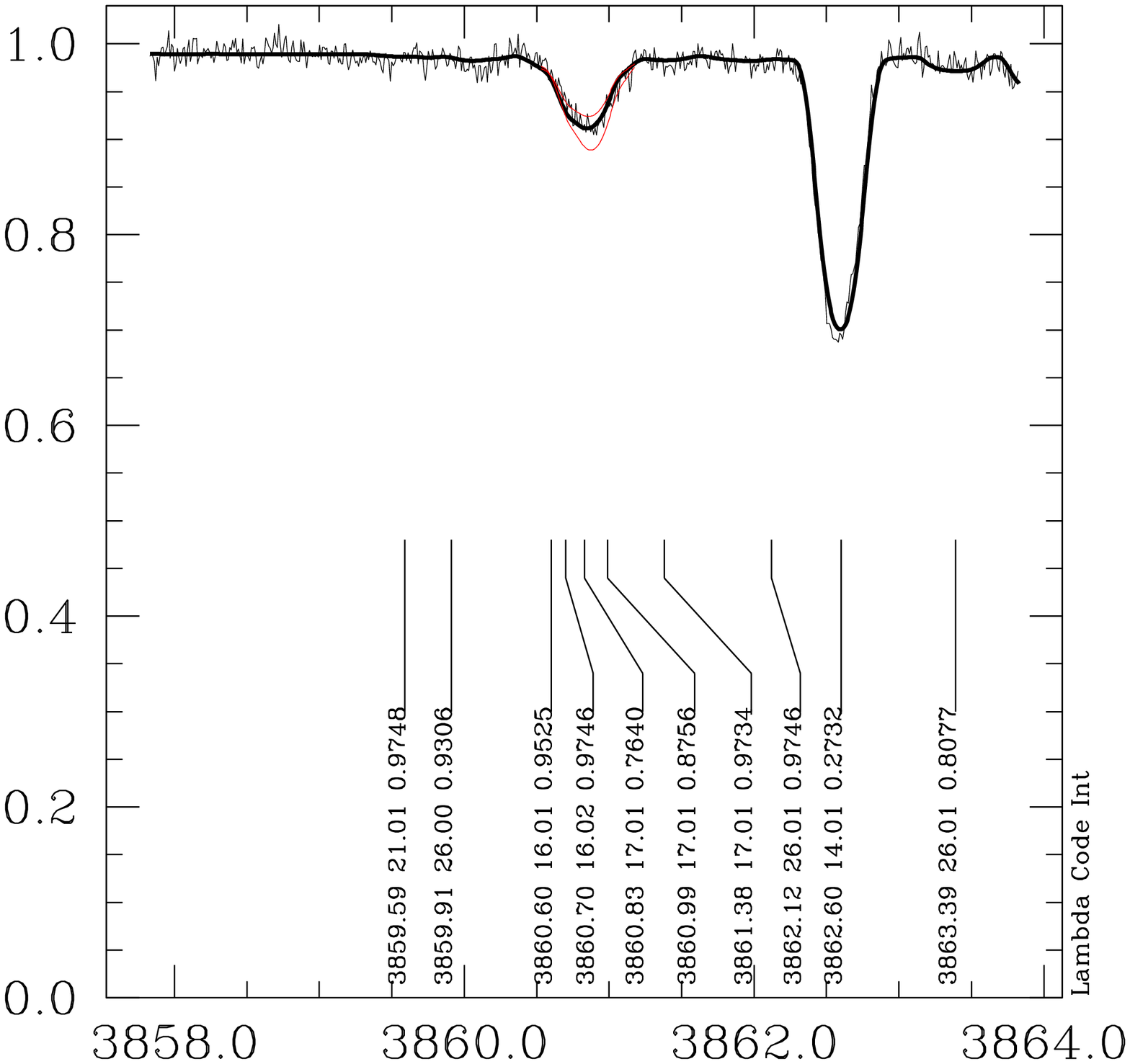}
\includegraphics[width=8cm]{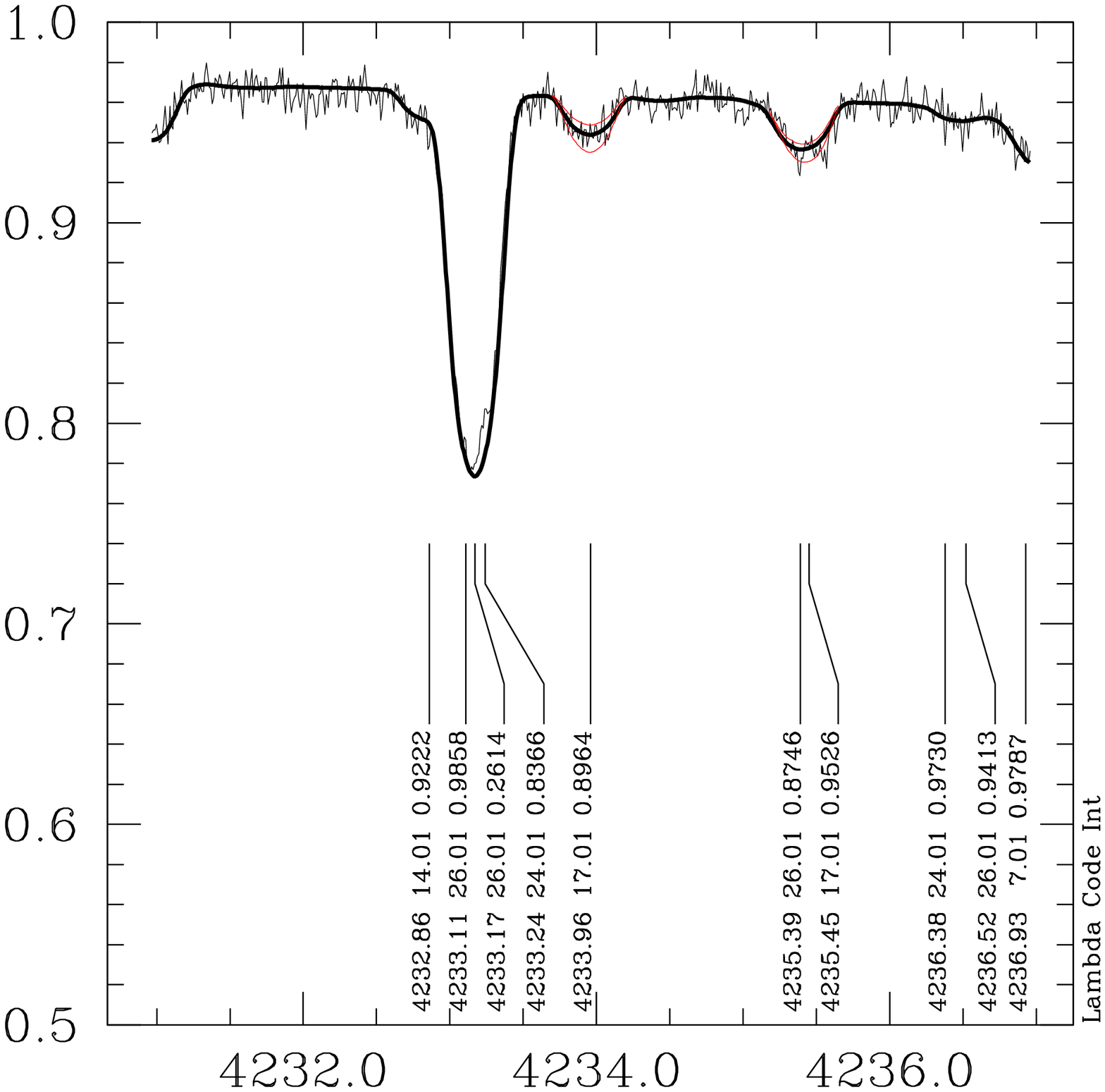}
\caption{Spectral region near the Cl II lines 3860 and 4233 (upper and lower
panels, respectively) identified in HD 5737. The observed and synthetic
spectra are shown by thin and thick lines, respectively.
The 2 thin red curves correspond to $\pm$1$\sigma$ abundance variation 
of the synthethic spectra. }
\label{clii.fig}
\end{figure}

HD 5737:  Most of their abundance values agree with the pattern of a He-weak star
(see Figure \ref{range.fig1}), with He I clearly underabundant ($\sim$0.7 dex below solar), 
and with intense Ti II, Cr II, Fe II, and Sr II. 
We also identified Cl II in this star and determined the abundance using eight 
different lines. Lines of Cl II that are clearly identified are
4233.9 {\AA} and the doublet 5217.74 {\AA}, 5217.94 {\AA}. 
We derived for Cl II an overabundance higher than $\sim$1 dex with a dispersion of
0.22 dex. In Figure \ref{clii.fig} we present two examples of the Cl II lines
3860.8 {\AA} and 4233.9 {\AA}, together with the synthetic fits to the observed spectra.

HD 21071: This is slightly subsolar but otherwise normal star.
For instance, the He and Fe abundances are -0.18 and -0.20 dex, respectively.
This star shows no patterns like any of the classical CP stars.

\begin{figure}
\centering
\includegraphics[width=8cm]{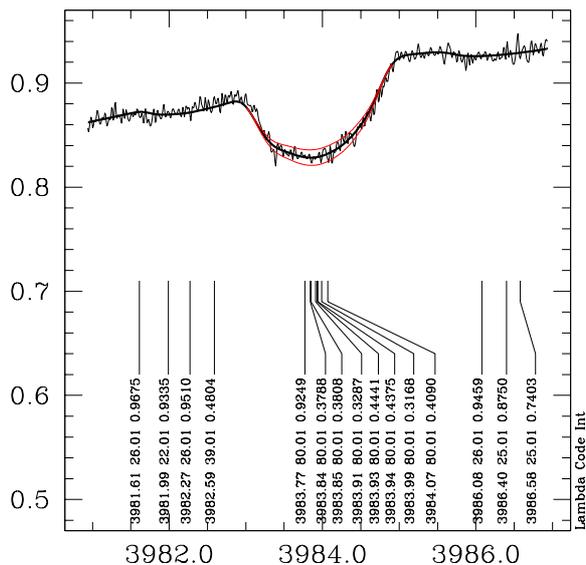}
\caption{Spectral region near the Hg II line 3984 identified in HD 23950.
The observed and synthetic spectra are shown by thin and thick lines, respectively.
The 2 thin red curves correspond to $\pm$1$\sigma$ abundance variation 
of the synthethic spectra. }
\label{hgii.fig}
\end{figure}

HD 23950: This is clearly a HgMn star (see Figure \ref{range.fig1}), with an excess
of both Mn II and Hg II relative to the Sun (2.16 dex and 5.18 dex, respectively). 
In Figure \ref{hgii.fig} we show the intense line Hg II 3984 identified
in HD 23950, together with their synthetic spectra.
The underabundances of C II (-0.44 dex), Mg II (-0.53 dex), and Si II (-0.34 dex),
together with overabundances of Ti II (0.81 dex), Cr II (0.52 dex), and Ga II (3.63 dex),
follow the general trend of the HgMn pattern.

HD 74146: This object has underabundances of Mg II (-0.41), Al II (-0.63), Si II
(-0.43), Ca II (-0.77), and Fe II (-0.68) compared to the Sun.
The subsolar value of Fe and other metals agree with a $\lambda$ Boo or a 
mild-$\lambda$ Boo star (see Figure \ref{range.fig1}).

HD 162586: Its abundances appear to be quite close to the solar values.  
This star shows no patterns like any of the CP stars.


HD 162630: This star presents solar or subsolar He (-0.16 dex relative to the Sun),
with slight metallic overabundances of Cr II (0.14 dex) and Fe II (0.13 dex).
The general pattern is similar to an Am or mild-Am star (see Figure \ref{range.fig2}),
including the overabundances of Al II (0.31 dex), Ni II (0.38 dex), and Sr II (0.75 dex).

HD 162678: This object present a slight deficiency of C II (-0.38 dex) and
Sc II (-0.29 dex), however most elements show almost solar abundances (Mg II, Al II,
Si II, Ca II, Cr II, and Fe II). This object does not follow the patterns of the
classical CP stars, and present nearly solar abundances.

HD 162679: Its abundances appear close to the solar values, except probably 
for Sc II (-0.45 dex). This object shows no patterns like any of the CP stars.

HD 162817: This object shows a subsolar He abundance (-0.36 dex relative to solar);
however, its temperature (10428 K) is lower than the usual for He-weak stars with
B spectral types (13000 - 15000 K). {{The common underabundaces of Ca and/or Sc
expected in Am stars, or the intense lines of Hg and/or Mn in HgMn stars,
are not observed in this object.}}
HD 162817 presents overabundances of Ca II (0.55 dex relative to solar), Sc II
(0.19 dex), Ti II (0.67 dex), Cr II (0.24 dex), and Fe II (0.29 dex), similar to an
Ap or mild-Ap (see Figure \ref{range.fig2}) star.

Our sample includes He-weak stars (HD 5737), Am or mild-Am stars
(HD 162630), mild-Ap stars (HD 162817), HgMn stars (HD 23950), 
mild-$\lambda$ Boo stars (HD 74146), and stars with abundances close to solar
(HD 21071, HD 162586, HD 162678, and HD 162679).
Within a temperature range of 10300 K - 14500 K, $\sim$40$\%$ of the sn stars
resulted CP stars. This value should be taken with caution due to how few
stars are studied here. However, the sn stars display a variety of chemical
peculiarities. No clear common spectral feature or abundance pattern is apparent,
beyond the sn characteristics itself. From the point of view of the chemical 
composition of the atmospheres, the sn stars are a rather inhomogeneous group
of stars. The sn characteristics appear in stars with very different chemical
properties.

We studied a sample of sn stars and not all of them display CP abundances,
i.e. some sn stars present $\sim$solar abundance values.
There is a lack of a clear relationship between sn stars and CP stars.
The anomalous chemical pattern of the CP stars is usually explained by the diffusion
processes. A supposed relation sn-CP would indicate that the sn characteristics are present
in those atmospheres with the physical conditions required by the diffusion to work efficiently.
This is also supported by their low rotational velocities.
However, the observational evidence shows that this does not seem to be the case.
We show that the sn characteristics can coexist simultaneously with the abnormal abundances,
but the chemical peculiarity is not a requirement for the sn phenomenon to appear.

Some stars in our sample agree with the general abundance pattern of the CP stars.
However, they also show some discrepancies compared to the general pattern.
A detailed inspection of classical CP stars shows that it is not unusual to find
some star-to-star differences in the abundances for each element.
Most of the chemical anomalies observed in the CP stars are usually explained by
the radiative diffusion \citep[e.g. ][]{michaud70,michaud76}, which is expected to depend
mainly on the effective temperature. However, the abundances could be significatively
affected, for instance, by the presence of magnetic fields \citep[see e.g. ][]{michaud81}.
There are other physical processes that could work in addition to the diffusion modifying
the abundances, such as mass loss \citep[e.g. ][]{vick08,vick10} or turbulence mixing
\citep[e.g. ][]{richer00,richard01}.
However, it is difficult to determine which of these mechanisms could produce the
resulting abundances. For instance, the abundances of Sirius A and of the Hyades star
68 Tau could be explained equally well by either mass loss or turbulence in addition
to diffusion \citep{richer00}.

Even for the case of the superficially "normal" stars, the elemental abundances of
late-B to F stars show a range of possible abundances \citep[][]{adelman-unsu07,hempel-h03}.
\citet{adelman-unsu07} found that the abundance distribution in normal A stars overlap
with those of the Am stars. Most normal A stars are likely to have a few values that
are of order $\pm$0.40 dex from solar, and this indicates that a common phenomenon could
produce the anomalies, such as a combination of diffusion and mass loss.
It is not uncommon to find stars considered "normal" with no exactly solar
abundance values. In our sample of sn stars, three objects display abundances within this
approximate range (HD 21071, HD 162586, and HD 162678).
\citet{hempel-h03} studied 27 B stars classified as normal and detected anomalous rather
than solar abundances in different elements (O, Mg, Ca, Fe, Sr, Ba). They show that the
anomalies do not occur to the same extent in all stars but instead show star-to-star variations.
The authors suggest that a counterpart to diffusion such as meridional mixing is operating in
the atmospheres of these normal stars.

\subsection{Other properties of sn stars}

We found no clear relation between the fundamental parameters
(T$_{\rm eff}$, log g and age) and the photospheric chemical composition of the
sn stars. This is probably related to a low statistical significance (only nine stars).
On the other hand, our sample includes five stars that belong to the same cluster (NGC 6475).
These five stars (HD 162586, HD 162630, HD 162678, HD 162679, and HD 162817)
allow a comparison where the age and the initial composition effects
are negligible. The group include two non-sn stars (HD 162630 and HD 162817)
and the other three are sn stars.
The range of temperature of the three sn stars is between 10219 K - 12500 K i.e.
mixed with the temperatures of the non-sn stars (10428 K and 10625 K).
The temperature alone does not seem to determine the sn characteristics.

The rotational velocities of the three sn stars in NGC 6475
(27 km/s, 40 km/s, and 40 km/s) are lower than the vsini values of the
two non-sn stars of the same cluster (49 km/s and 75 km/s).
The apparent preference of the sn stars for objects with low vsini
agrees with the result of \citet{mermilliod83}. However, this should
be taken with caution owing to the low number of objects studied here.

\citet{mouj98} studied the spectrophotometric variations of 49 Be and shell-like stars.
They found six with relatively low vsini values that are candidates to exhibiting 
"spectrophotometric shell behaviour". These spectrophotometric criteria should be taken
with caution because the presence of absorption lines is required for a shell star.
However, none of the six stars proposed by \citet{mouj98} present evidence of shell
absorption lines \citep{rivinius06}. 
On the other hand, as mentioned in the Introduction, \citet{neiner05}
identified a group of Be and shell stars in the field of view of the COROT satellite.
In particular, one of the stars they discovered (HD 174512) is classified either as a
slowly rotating (vsini$\sim$20 km/s) shell star or as a Herbig Be star.
If the shell-like nature of this object is confirmed, this could be the first star of
this class with a low projected rotational velocity.

For the case of CP stars, it is not fully understood how the abundances are affected by
stellar rotation. Once the He convection zone has disappeared in stars with vsini lower than
75 km/s \citep{charb-michaud88}, stable atmospheres and diffusion are expected to occur.
\citet{charb-michaud91} carried out detailed calculations of
diffusion in the presence of meridional circulation and turbulence. They conclude that
no dependence of abundance anomalies on vsini is expected for most elements,
because diffusion time scales are much shorter than a meridional circulation time scale.
However, the authors note that their calculations do not imply that there
should be no correlation at all between rotation and abundance anomalies. 
Some effects, such as the dependence of the He cutoff velocity on log g and on
stellar evolution, could lead to a weak anticorrelation between abundances and rotation.
\citet{burkhart79} did not find any decrease in abundance anomalies
with increasing vsini. On the other hand, \citet{boyarchuk-savanov86} show that
the Vanadium overabundance decrease with increasing rotational rate.
The lack of correlation between abundances and rotation show that the meridional circulation
has little influence on chemical separation for vsini$<$100 km/s \citep{charb-michaud91}.

\subsection{Line profiles of sn stars}

\citet{abt-levato77} mentioned that their 14 sn stars have both sharp and
broad lines of He I, along with sharp metallic lines. 
Another example is presented in Figure 2 of \citet{abt78}, where the
sn star HD 36392 presents both broad lines of He I (4009 \AA, 4026 \AA, 4144 \AA)
and sharp "normal" lines of He I (4121 \AA). 
In other words, sn stars usually present a sharp lined
spectra, except for some particular He I lines (4009 \AA, 4026 \AA, 4144 \AA).
However, in our sample the line 4009 {\AA} is weakly present or not seen in most of
the stars, while the line 4026 {\AA} is always present, since it is probably the clearest example
of a broad He I line in the sn stars of our sample.
It is notable that we have identified the sn characteristic in our group with
only one broad He I line in common.
This may give rise to the question about whether the sn stars are unusual
in their He lines compared to other normal stars.
This topic is addressed in this section.

We explore three different physical proccesses, in order to determine whether they
have some contribution (or not) to the particular line profiles observed in the sn stars.
The processes discussed include NLTE effects, broadening mechanisms of spectral lines,
and the possible stratification of the atmospheres. In particular, we focus on the broad
He I lines that are present in the sn spectra.

\subsubsection{NLTE effects in the He I lines}

\begin{figure*}
\centering
\includegraphics[width=8cm]{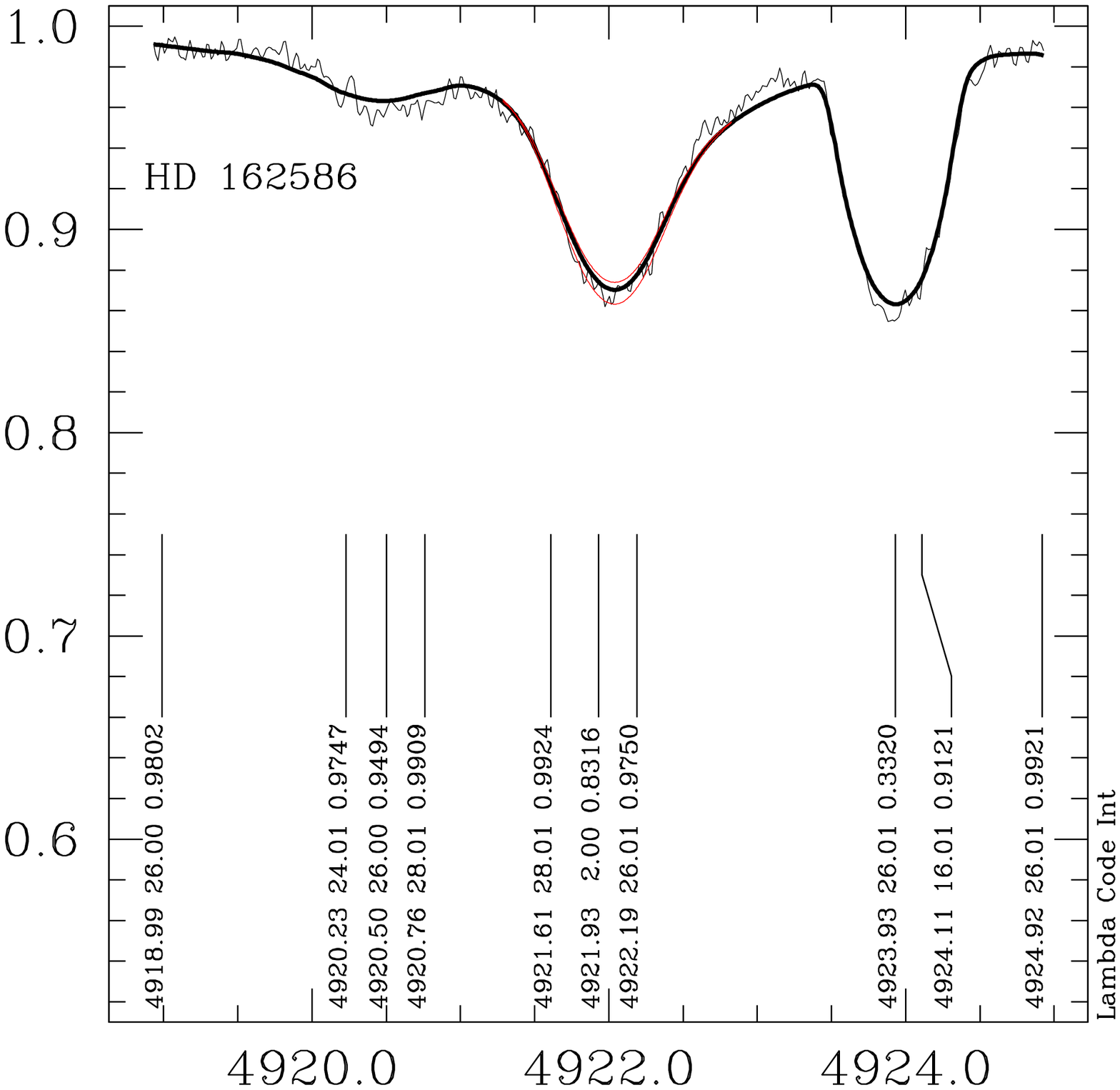}
\includegraphics[width=8cm]{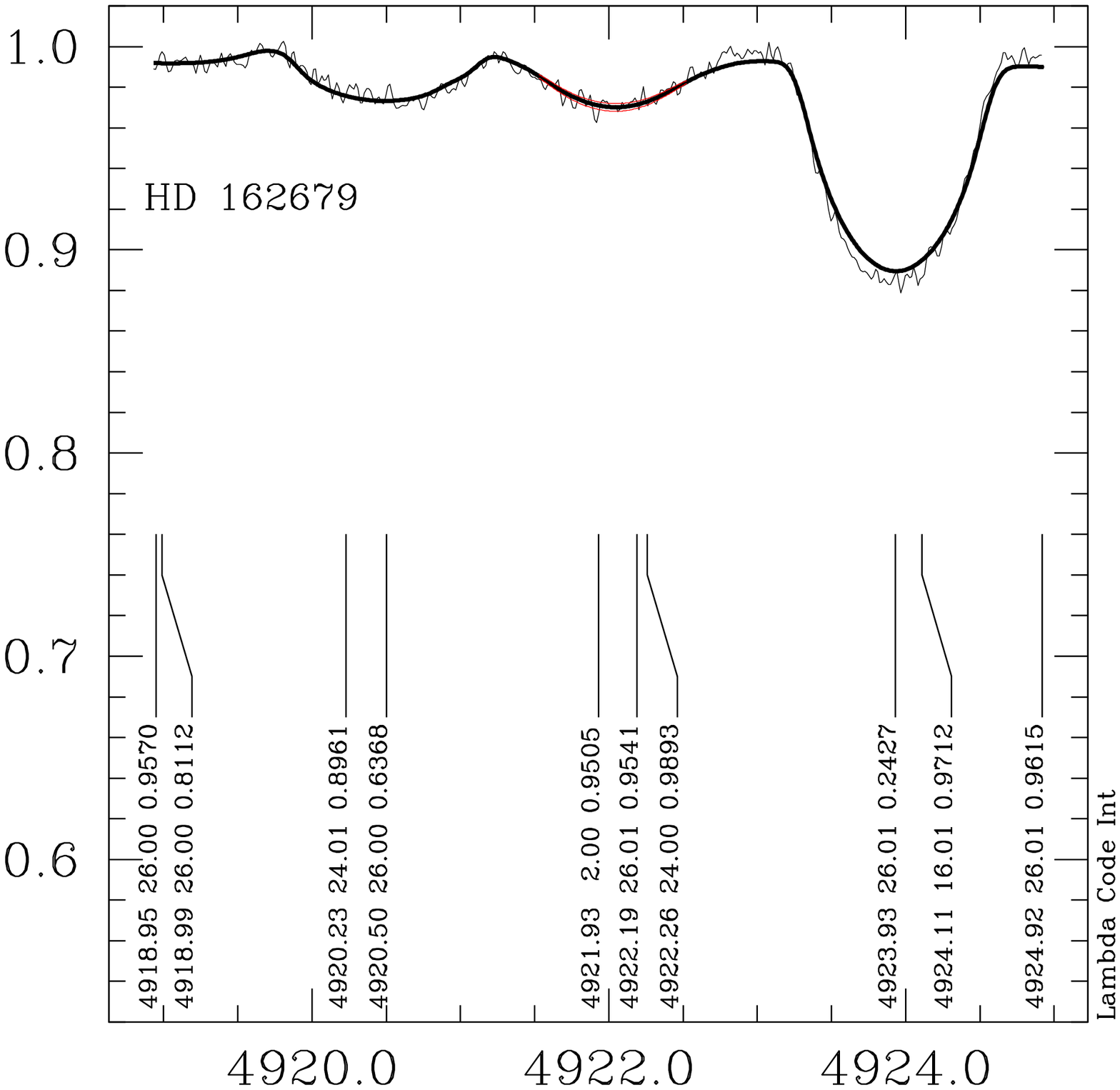}
\includegraphics[width=8cm]{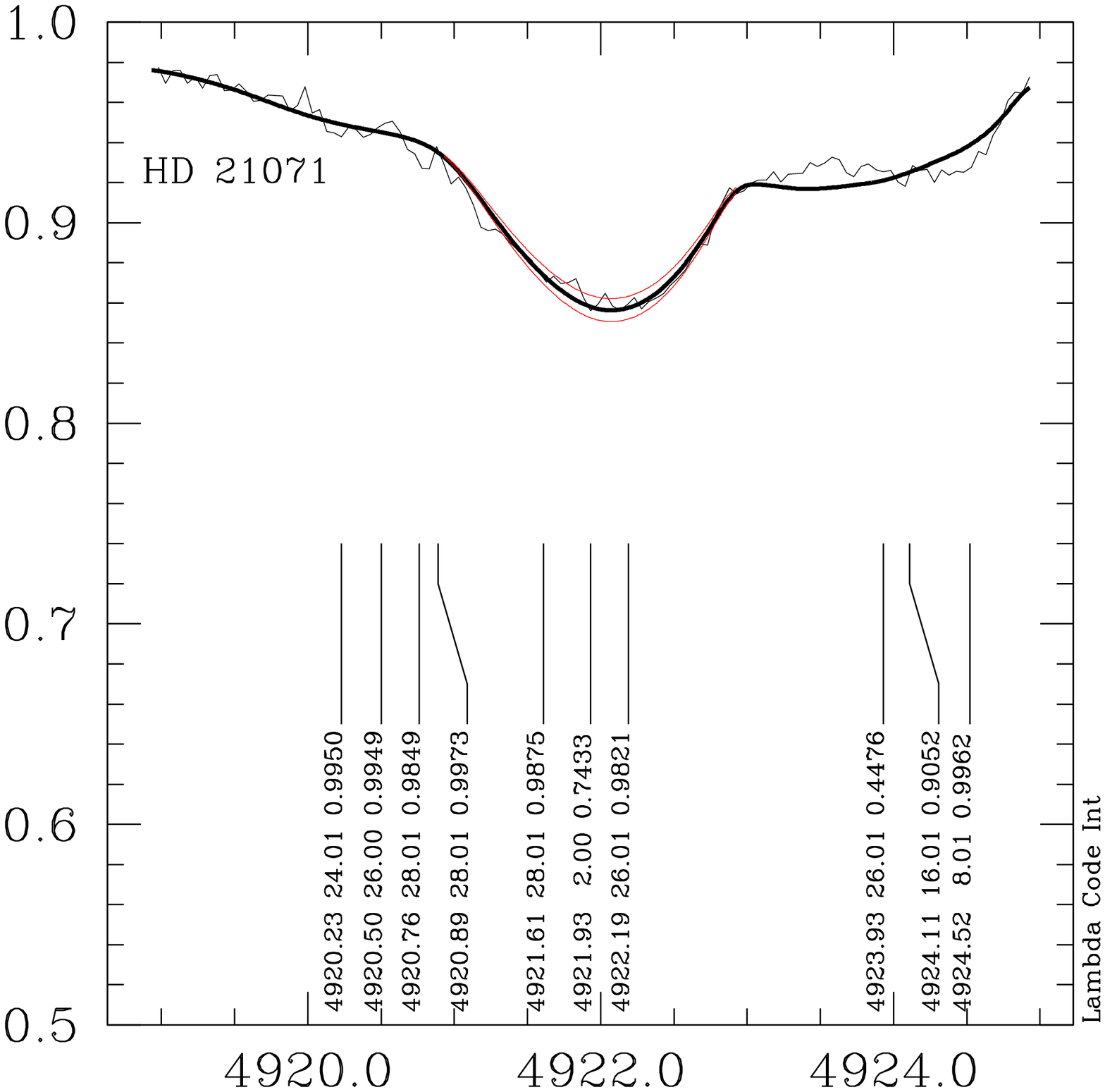}
\includegraphics[width=8cm]{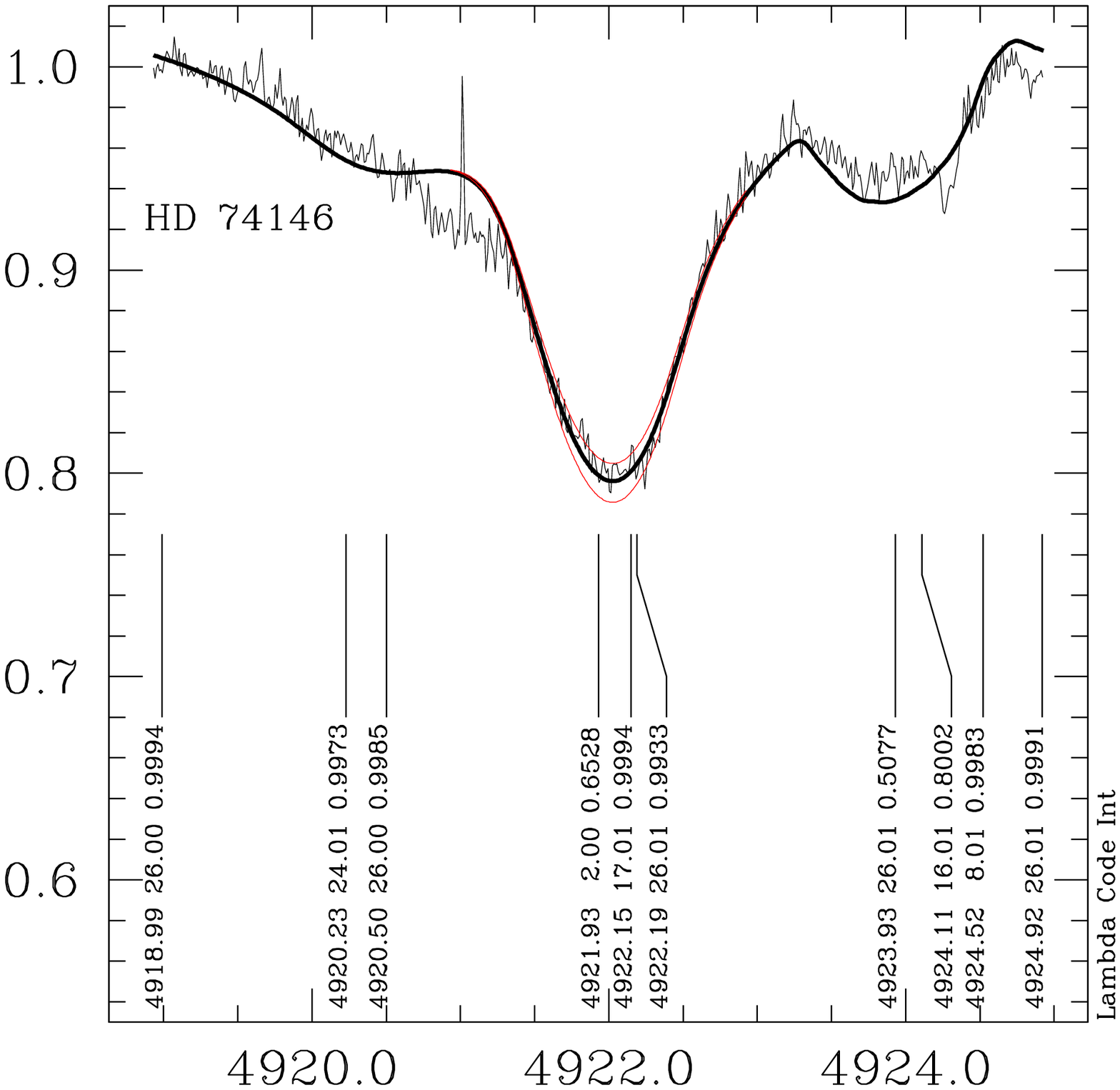}
\caption{Spectral region near the He I line 4921 {\AA}.
The observed and synthetic spectra are shown by thin and thick lines,
respectively. The panels correspond to the stars HD 162586, HD 162679, HD 21071, and HD 74146.
The 2 thin red curves correspond to $\pm$1$\sigma$ abundance variation 
of the synthethic spectra. }
\label{fig.nlte.4921}
\end{figure*}

Different lines of He I are more or less sensitive to the NLTE effects.
To verify possible presence NLTE effects in the He I lines, we compared in
Fig. \ref{fig.nlte.4921} the observed vs LTE synthetic spectra for the line 4921 {\AA}, which
is one of the most sensitive He I lines to the NLTE effects \citep[e.g. ][]{przybilla11}.
The example corresponds to the sn stars HD 162586, HD 162679, HD 21071, and HD 74146, where
this line is intense, and then the NLTE effects should be noted first.
The agreement between observed and LTE synthethic spectra for this line
shows that NLTE effects are not clearly detected in the He I profiles
of these stars.

If NLTE effects are present in the profiles of He I lines, they are mainly
noted as an enhancement in the cores of lines \citep[e.g. ][]{przybilla11}.
The cores of the lines are formed in the higher layers of the atmosphere
where the density is lower, the number of interactions between particles decrease,
and the radiation field tends to dominate.
However, we have observed a broadening in the wings of He I lines such as 4026 {\AA}
in our sample rather than an enhancement in the cores.
In other words, NLTE effects and the sn characteristics involve, in principle,
different parts of the He I profiles (cores and wings, respectively).
Also, NLTE effects are first noted in their most sensitive lines, such as
4921 {\AA}, 5875 {\AA}, and 6678 {\AA} (e.g. Przybilla et al. 2011), while the sn
characteristics are identified first in other lines such as 4026 {\AA} in our sample.
Although NLTE effects could not be completely ruled out, it does not
seem that the sn characteristics are directly related to the NLTE effects.

\subsubsection{Broadening of He I lines}

The intrinsic line profiles in the synthetic spectra take 
the effects of natural, Stark, Van der Waals, and thermal Doppler broadening into account.
The computed spectra must be convolved with the instrumental profile
and broadened by the star rotation.
For temperatures higher than around 10000 K, the H is mainly ionized, and
the dominant line broadening mechanism in B stars is the quadratic Stark effect,
owing to interactions of absorbing atoms with charged particles \citep[e.g. ][]{gray}.
There are different methods of computing the Stark broadening $\Gamma$ function
of the spectral lines \citep[e.g. ][]{cowley71,kurucz74,popovic01}.
\citet{kurucz74} and \citet{cowley71} describe two separate methods
of calculating the Stark broadening $\Gamma$ independent of temperature, while the \citet{popovic01}
method scales the \citet{kurucz74} broadening by the square root of the temperature.
For most of the spectral lines, the ATLAS9 and SYNTHE codes use the \citet{kurucz74}
method for calculating the Stark broadening.

The He I lines could be incorrectly modelled if the Stark broadening is not adequately
taken into account \citep[e.g. ][]{przybilla11}.
To improve the modelling of the He I profiles, we used detailed line broadening tables
for the line 4471 {\AA} \citep{barnard74} and for the lines 4026 {\AA}, 4387 {\AA}, and 4922 {\AA}
\citep{shamey69}. Most of these profiles have been computed assuming quasistatic broadening
by ions and impact broadening by electrons, with the semiclassical broadening theory for
overlapping lines of \citet{griem62}. Although these detailed calculations do not include all
lines of He I, they do include the line 4026 {\AA}, which clearly displays the sn characteristics
(a broad He I line) in all stars of our sample.

\begin{figure*}
\centering
\includegraphics[width=8cm]{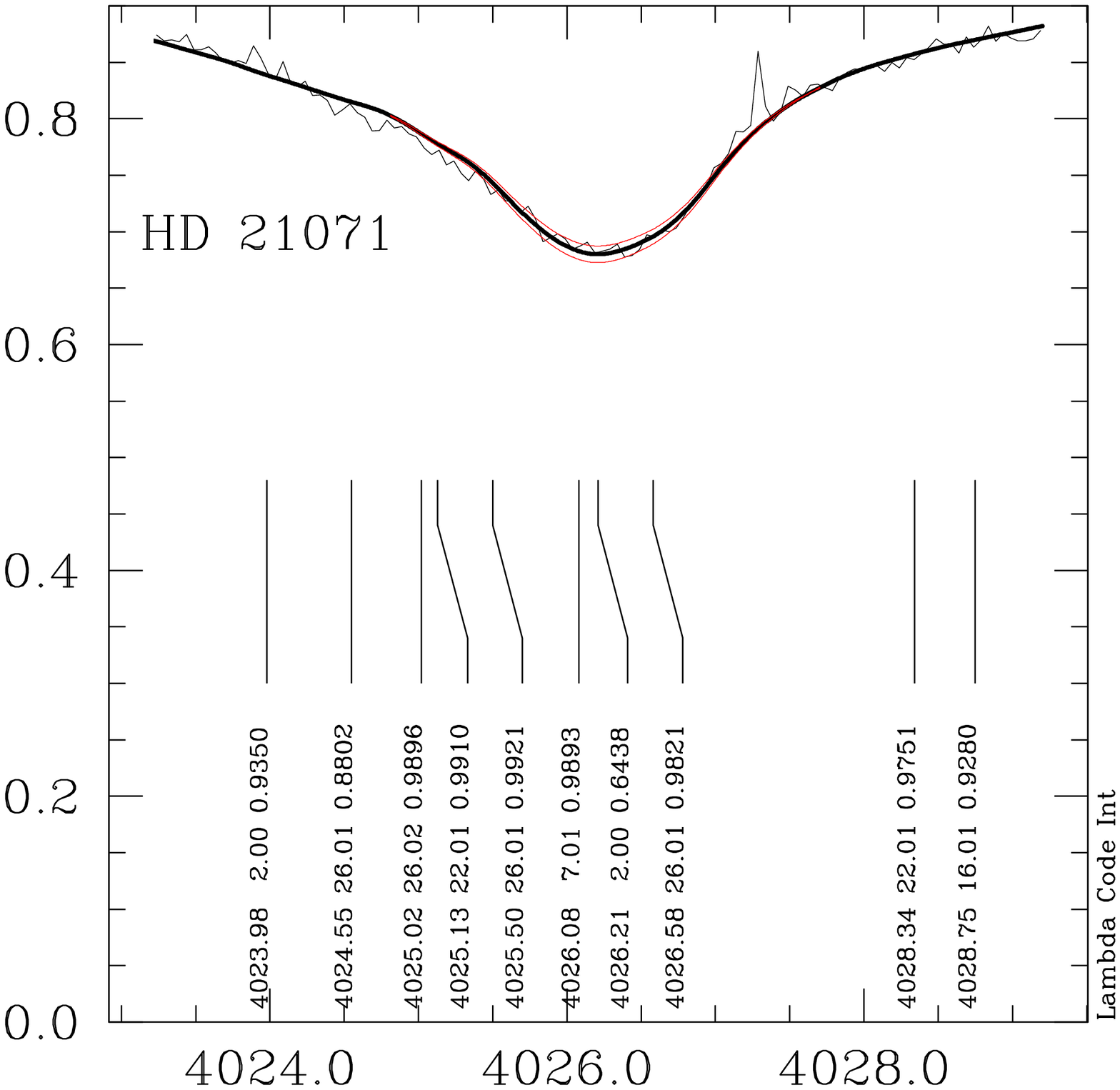}
\includegraphics[width=8cm]{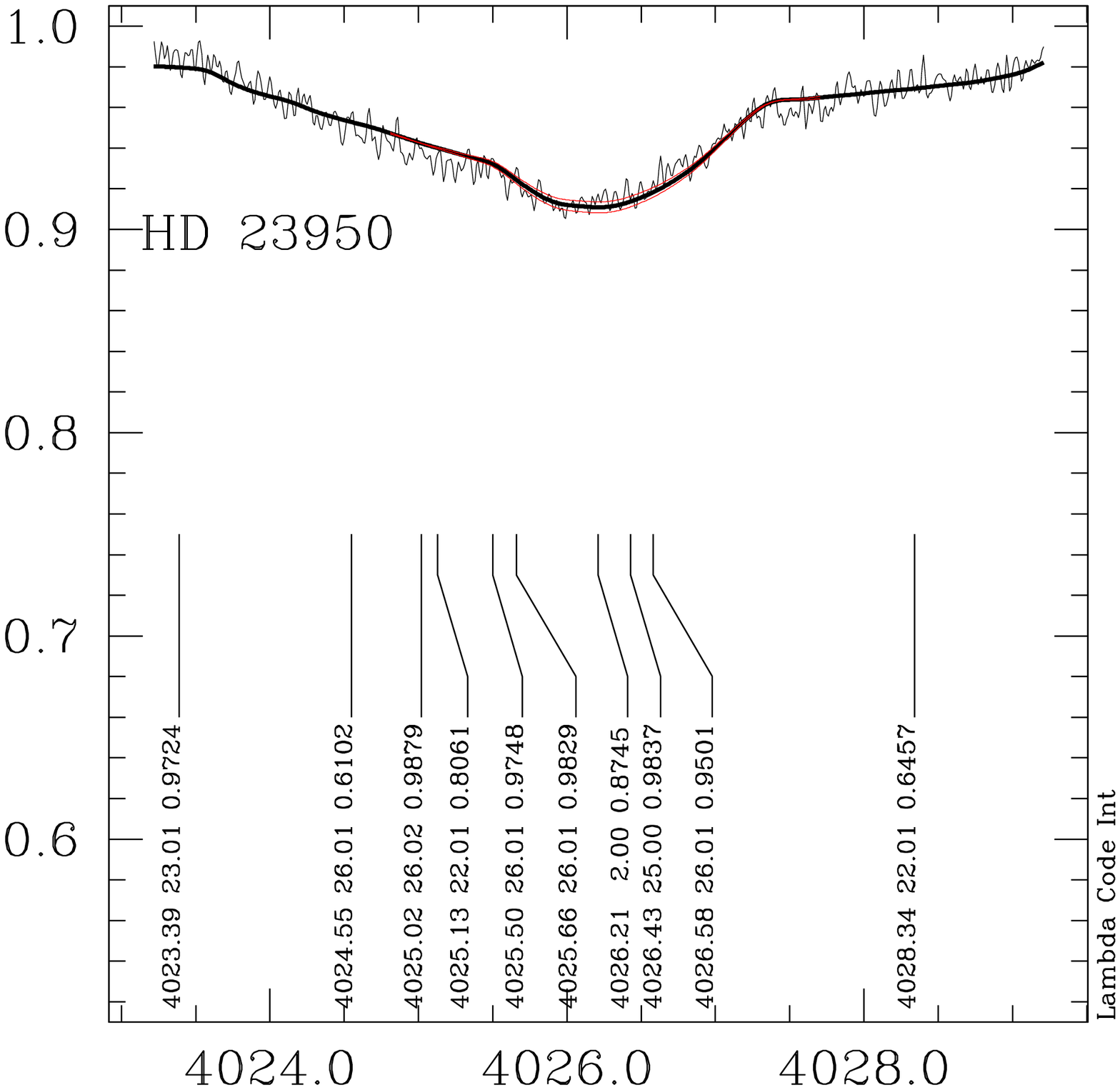}
\includegraphics[width=8cm]{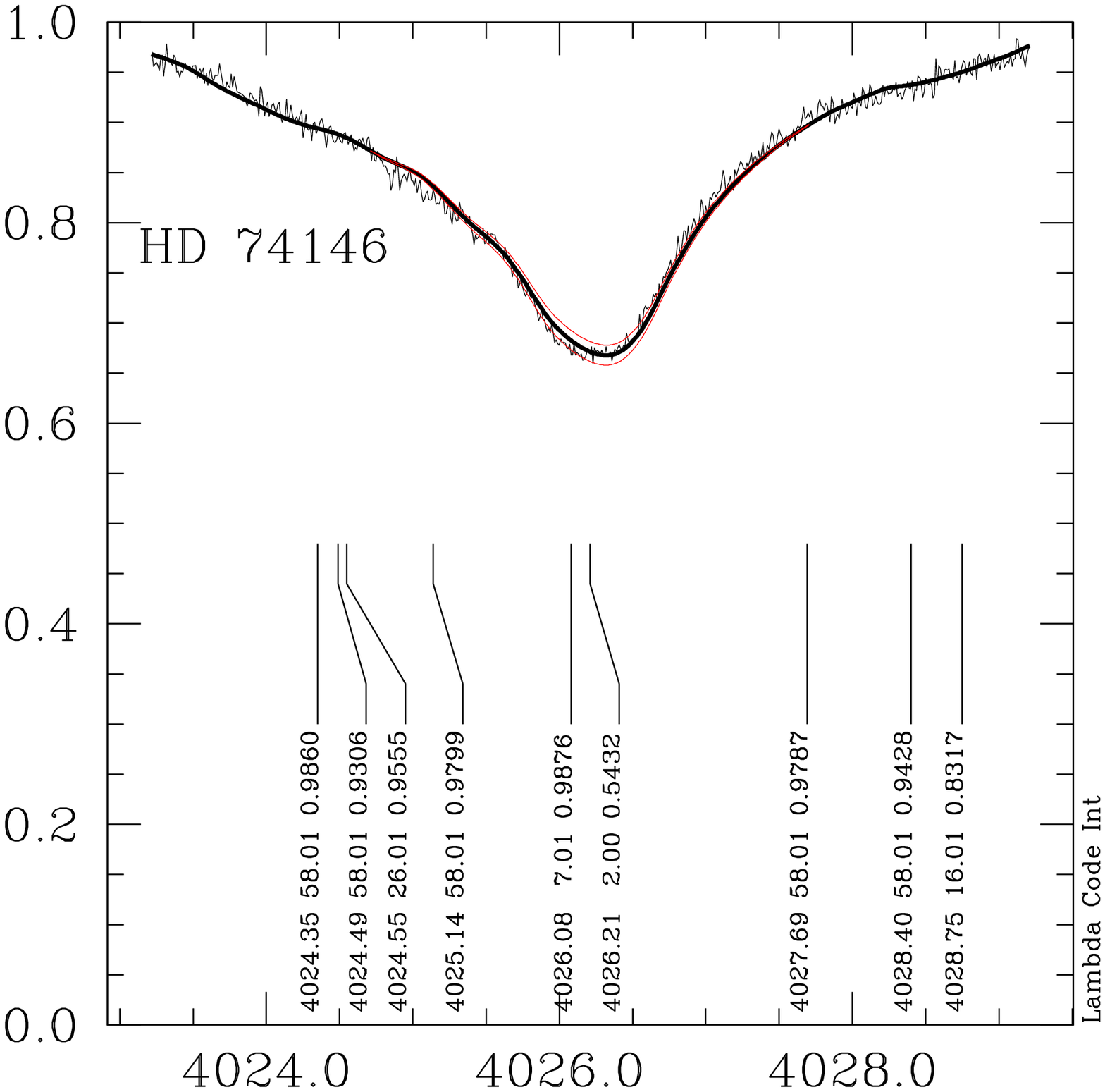}
\includegraphics[width=8cm]{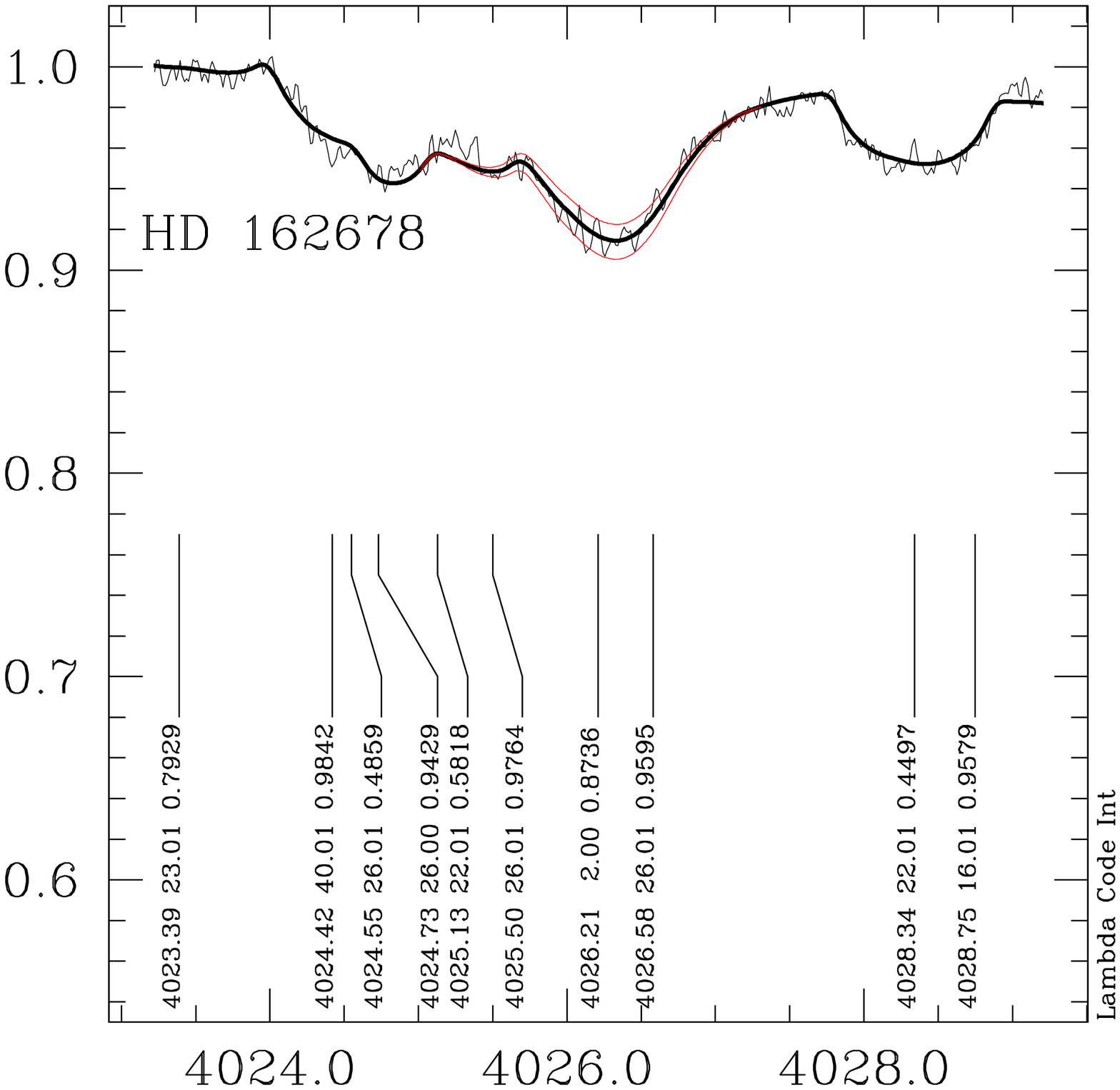}
\caption{Spectral region near the He I line 4026 {\AA} for a sample of sn stars.
The observed and synthetic spectra are shown by thin and thick lines,
respectively. The panels correspond to the stars HD 21071, HD 23950, HD 74146 and HD 162678.
The 2 thin red curves correspond to $\pm$1$\sigma$ abundance variation 
of the synthethic spectra.  }
\label{fig.4026.a}
\end{figure*}

\begin{figure*}
\centering
\includegraphics[width=8cm]{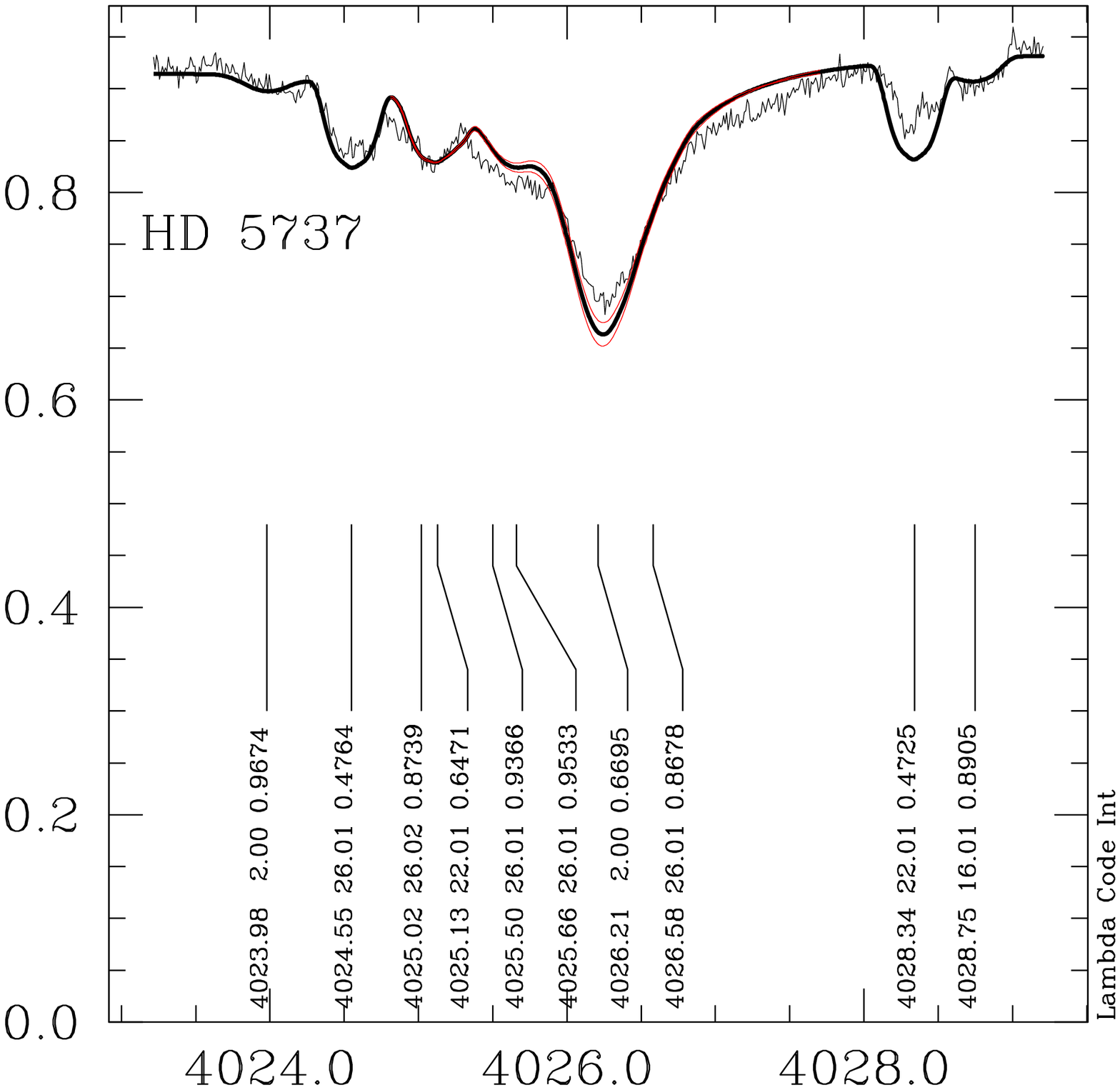}
\includegraphics[width=8cm]{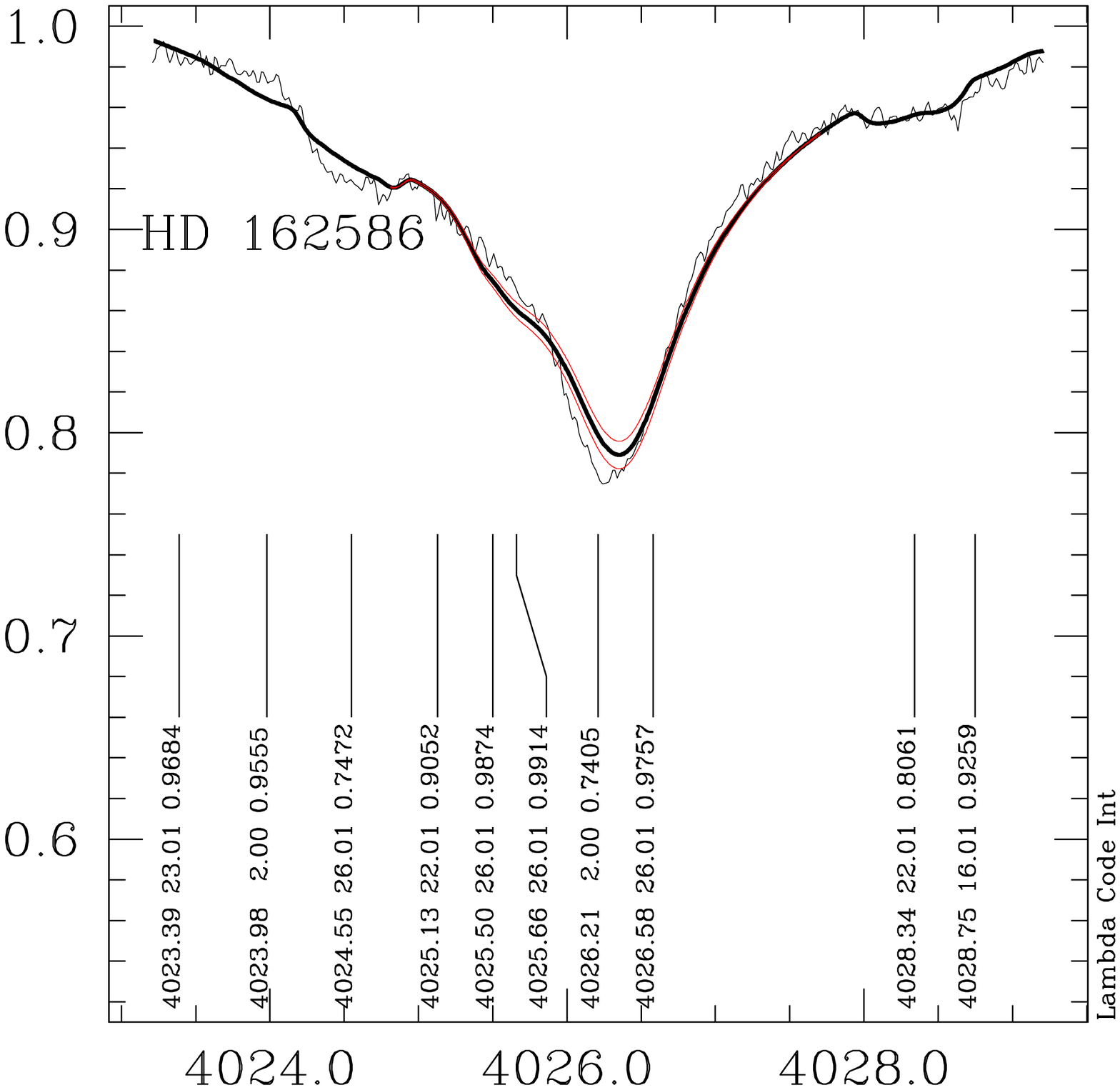}
\includegraphics[width=8cm]{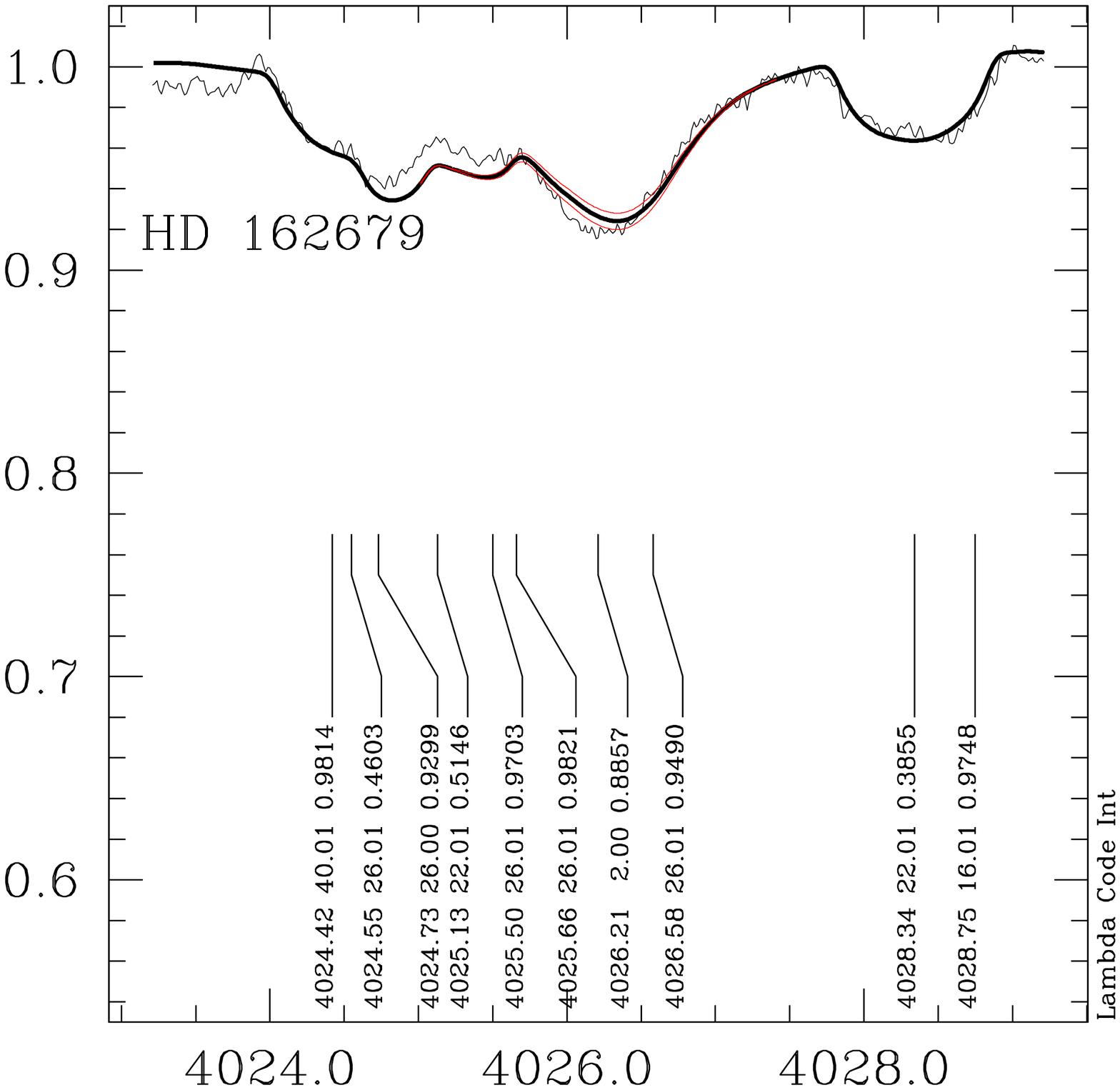}
\caption{ Spectral region near the He I line 4026 {\AA} for a sample of sn stars.
The observed and synthetic spectra are shown by thin and thick lines,
respectively. The panels correspond to the stars HD 5737, HD 162586 and HD 162679 for which
the synthethic spectra does not perfectly fit the observations.
The 2 thin red curves correspond to $\pm$1$\sigma$ abundance variation 
of the synthethic spectra. }
\label{fig.4026.b}
\end{figure*}

The Figures \ref{fig.4026.a} and \ref{fig.4026.b} show the observed vs synthethic spectra
derived for the line 4026 {\AA}, including the detailed Stark broadening tables in the calculation.
We mentioned that the observed broadening in this line is the clearest example of the sn
characteristics (broad He I line) in our sample. 
We note small differences between observed and synthethic spectra for the stars HD 5737, HD 162586, and
HD 162679 (see Figure \ref{fig.4026.b}), although (even for these stars) the general broadening
seems to agree with observations.
The small differences in these three stars do not seem to be NLTE effects, because this is not
detected in the more sensitive He I line 4921 {\AA} (see next section for another possible effects).
We succesfully fitted the He I 4026 {\AA} in most sn stars (4 out of 7) by adopting the correct
Stark broadening tables, including the instrumental and rotational convolutions. This modelling
seems to be enough to describe the profile of the line 4026 {\AA}, without needing to include
an extra broadening mechanism, except probably for HD 5737, HD 162586, and HD 162679.

There is good agreement between the He I abundance values derived from the broad
lines (4009 {\AA}, 4026 {\AA}, 4144 {\AA}) and from the rest of the He I lines.
There is no clear tendency between them in the sn stars, i.e., the abundances obtained
from the broad He I lines are not greater or lower than the sharp He I lines.

\subsubsection{Search for stratification of He I lines}

Studying He-peculiar stars in particular, \citet{vauclair91} predict that the He
abundance should decrease with depth in the atmosphere of He-rich
stars and increase in He-weak stars; i.e., both should present He stratified atmospheres.
\citet{farthmann94} proposed that the excesively broad wings of the He I line 4471 {\AA}
observed in the He-weak star HD 49333 are due to a vertical He stratification.
They require the use of stratified instead of homogeneous models to account for the observed
wings in this line, by adopting a greater He abundance in the deepest layers.
\citet{leone-lanzafame97} confirm the Vauclair et al. result for He-rich stars; 
however, they found the contrary effect for He-weak stars (i.e. the abundance decrease with depth)
by studying the He I red spectral transitions 6678 {\AA}, 7065 {\AA}, and 7281 {\AA}.
This is also in contrast to \citet{farthmann94}, in which the abundance increases with depth
in the atmosphere.

More recently, \citet{zboril05} has found that 5 out of 19 He-rich stars are probably He-stratified, studying
the 4026 {\AA} line profiles. \citet{catanzaro08} finds that the He I abundance decrease towards the deepest
layers in the He-rich star HD 145792 by using a plot of He I abundance vs optical depth for different
lines. He also noted that the impossibility of fitting the line 4026 {\AA} with a single abundance value is
a sign of stratification. For this line, the author reproduce the core but not the red wing of the profile.
Although there are some discrepancies if the He increase or decrease with depth in He-peculiar
stars, the mentioned works agree that the He is possibly stratified and that this could affect the profiles
of 4471 {\AA} \citep{farthmann94} and 4026 {\AA} \citep{zboril05,catanzaro08}.

\citet{leone98} found that the He-rich star HD 37017 is He stratified in their magnetic pole regions.
This variable star show a stratified atmosphere (in a plot of abundance vs depth) when observed in the phase
of helium maximum strength, however when looking in the minimum helium phase, the star appears to be
unstratified. In other words, an analysis of a single observation does not seem to completely rule
out the possibility of stratification.

\begin{figure}
\centering
\includegraphics[width=6cm]{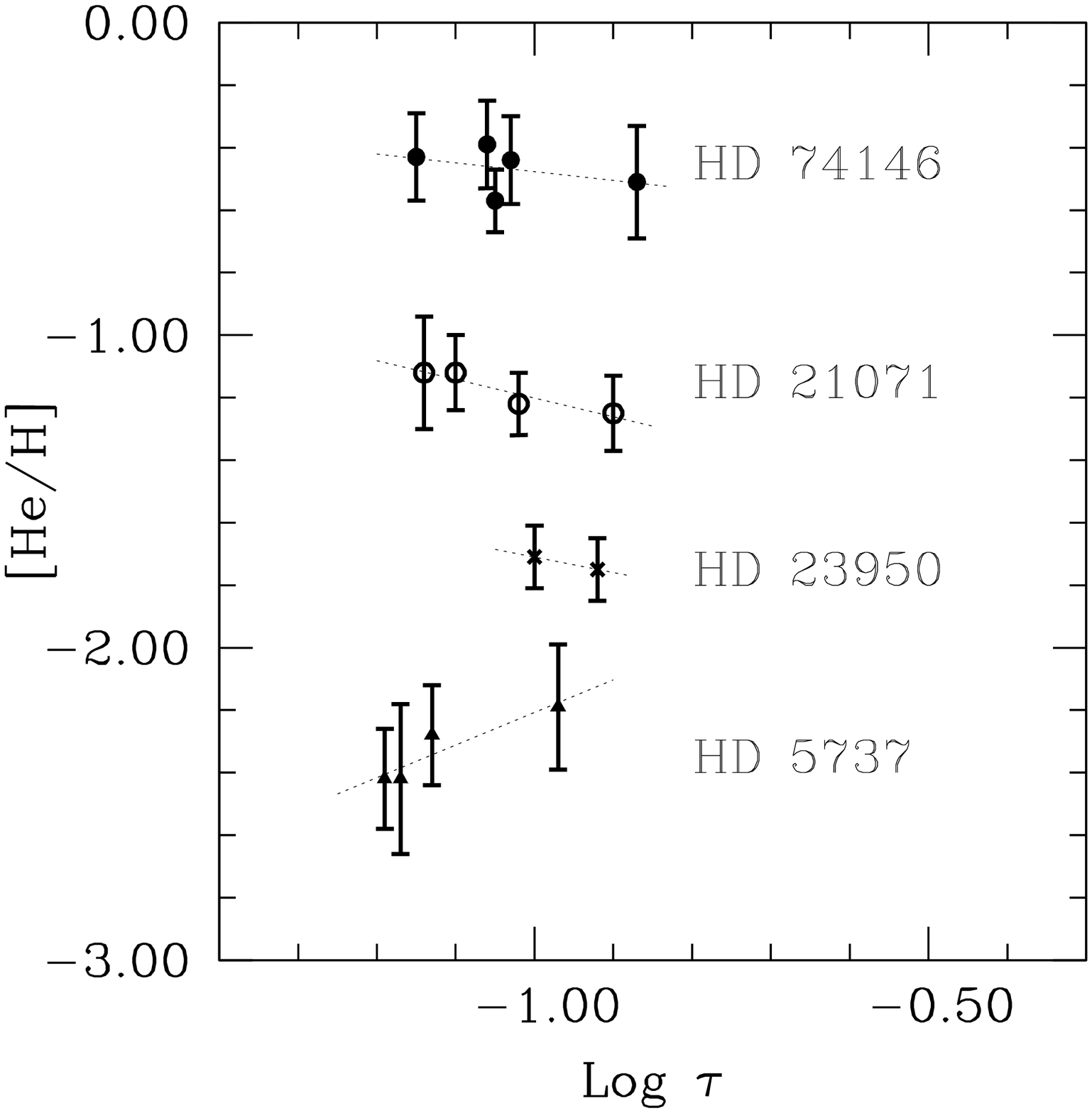}
\includegraphics[width=6cm]{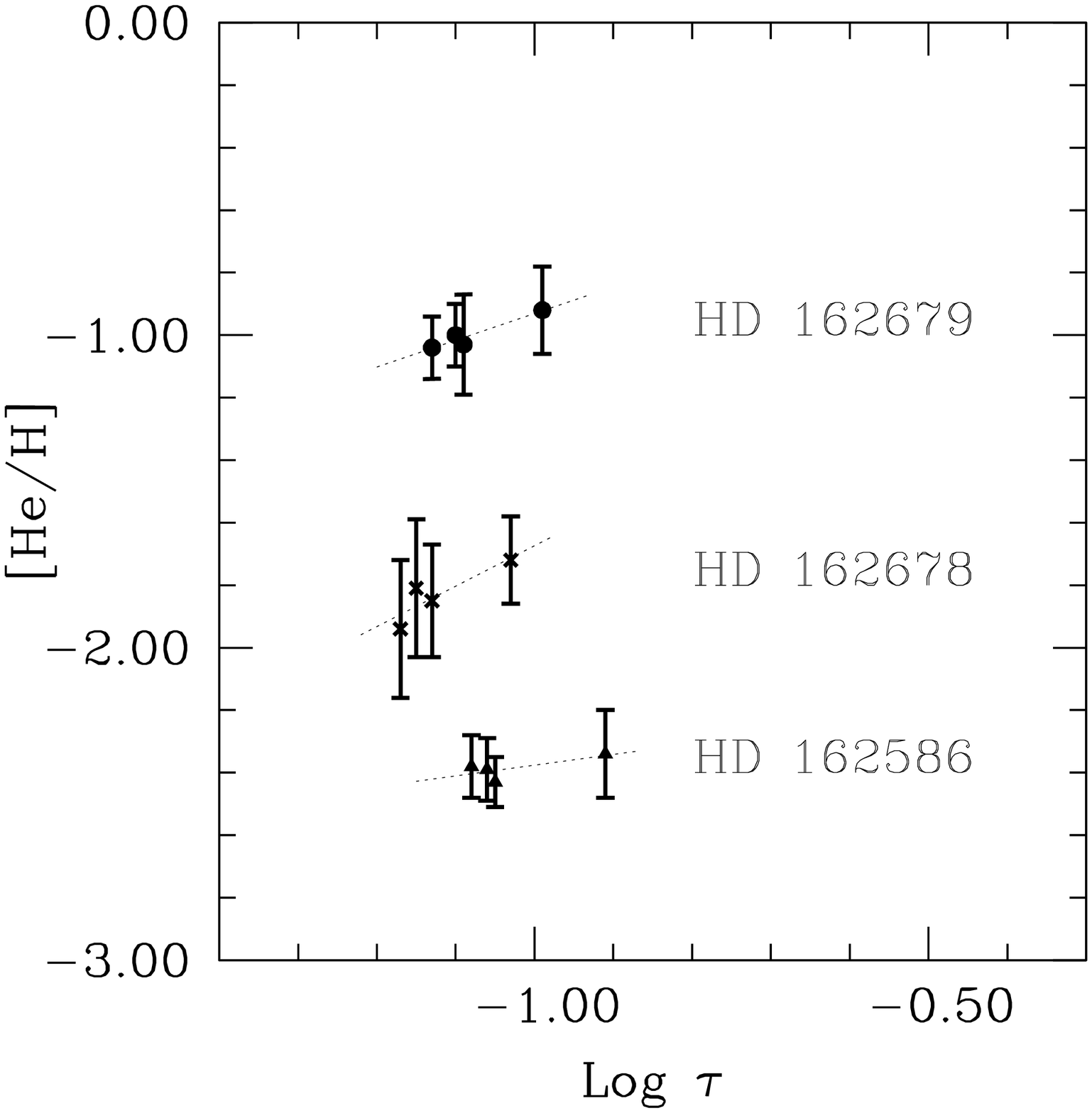}
\caption{Abundances of He I lines vs. optical depth for the sample of sn stars.
The y-values were displaced vertically to avoid the superposition of points.
The dotted line shows a linear fit to the data.
}
\label{fig.stratif}
\end{figure}

In Figure \ref{fig.stratif} we present a plot of abundance vs. optical depth for the He I lines
measured in the sn stars, including the error bars of the points.
The significance of the linear fit to the data
could be estimated by comparing
the slope m of the fits with their corresponding error $\sigma_{m}$.
The y values were displaced vertically in the figure to avoid the superposition of points.
The cores and wings of the lines are formed mainly in the higher and lower layers of the atmosphere,
respectively. Then, a possible wing broadening in the He I lines would correspond to an increase in the
He abundance with depth \citep[e.g. ][]{farthmann94}. From Figure \ref{fig.stratif}, this effect
seems be present in the sn stars HD 5737 (m$\sim$3.4$\sigma_{m}$), 
HD 162678 (m$\sim$2.6$\sigma_{m}$), and HD 162679 (m$\sim$4.6$\sigma_{m}$),
and less clearly in HD 162586 (m$\sim$1.5$\sigma_{m}$).

We caution that the low number of points presented in Figure \ref{fig.stratif}, together with
the error bars of the individual measurements, could diminish the significance of a possible trend.
However, these plots give a first indication of a possible stratification in these atmospheres
\citep[e.g. ][]{catanzaro08}.
We showed that HD 5737 is a He-weak star, so then a possible He-stratification could not be discarded.
The stars HD 162678 and HD 162679 present almost solar He abundance values, within $\sim$0.13 dex. 
From Figure \ref{fig.stratif} a possible increase in the He abundance with depth could
not be ruled out in the stars HD 5737, HD 162679, and less probably in HD 162586. For these stars,
a stratified model atmosphere probably improves the agreement with the observed profile of
the He I line 4026 {\AA}.

Figure \ref{fig.stratif} suggests that different abundance values
could correspond to different depths in the atmosphere.
If the abundance increases with depth, the cores of the individual lines
should be adjusted with a lower abundance value than for the wings.
However, it is difficult to see this effect in the profile of
a single line such as He I 4026 {\AA} (see Figure \ref{fig.4026.b}).
We note that the final abundance value adopted for these lines corresponds to the
best fit including both the core and the wings.
The curves of $\pm$1$\sigma$ abundance values are indicative of the
effect of increase or decrease in the abundance of the line.
A variation in the abundance is first noted in the core of the
profile, and then it is (progresively) less intense towards the wings.
For the stars HD 162586 and HD 162679, the blue wing of He I 4026 {\AA}
is not properly fitted (see Figure \ref{fig.4026.b}).
Many attempts were made, although without success. This is probably due
to strong blends with Fe II 4024.55 {\AA}, Fe III 4025.02 {\AA}, and
Ti II 4025.13 {\AA} for the case of HD 162679.
The red wings are better fitted although not perfectly.
The cores of He I 4026 {\AA} seem to be slightly blue-shifted in wavelength
only in these two sn stars, and cannot be properly fitted even for
different abundance values.
For the star HD 5737, we note that the red wing of He I 4026 {\AA} (roughly
between $\sim$4026.2 and $\sim$4028 {\AA}) could not be fitted by varying the abundance. 
The core of the line shows better agreement with a lower abundance
(which corresponds to an increase in abundance with depth), although
it does not fit perfectly. For the three stars of Figure \ref{fig.4026.b},
it is very difficult to fit the line He I 4026 {\AA}, even when adopting
different abundance values for the cores and wings.
This suggests that another effect is probably superimposed, such as a
non-uniform distribution of He in the surface \citep[e.g.][]{bohlender89}.

{{The He-strong sn star $\delta$ Ori C was studied by \citet{bohlender89} }}
using a high-resolution spectra taken at the CFHT. The author found that the He I line profiles
cannot be reproduced with a uniform atmospheric He abundance. He proposed two explanations:
a non-uniform distribution of He in the stellar surface or a vertical He stratified atmosphere.
This work shows that the He-stratified atmospheres cannot be ruled out in at least some sn stars
by using the plots of abundance vs. depth (see Figure \ref{fig.stratif}).
The non-uniform distribution of He could be analysed by taking spectra at different rotational
phases. The study presented here is mostly based on a single observation of the sn stars. 
A further study of variability in the He I lines would help determine if a possible
non-uniform He distribution could play a role in the sn stars.

In the previous sections, we used a model atmosphere with given physical conditions and appropriate
line broadening, and fitted both broad and sharp lines of four out of seven sn stars.
In particular, we properly calculated the broad He I line profiles that characterize and help
define the class of sn stars.
Therefore we showed that at least in some stars the broadening could have originated
in the atmospheres of these stars. There is no need to suppose another mechanism or an
extra-broadening source for the He lines. The usual radiative transfer processes that give rise
to the "normal" He line formation and include the Stark broadening seem to be enough to describe
the line spectra of sn stars.
For the three other stars, the small differences in the profiles seem to be due, at least in part,
to stratification (increase He abundance with depth) in these atmospheres, although
a non-uniform He surface distribution could not be ruled out. We suggest more observations and
the use of stratified model atmospheres to properly describe the He I profiles in these sn stars.

\section{Concluding remarks}

We have derived the chemical abundances of nine stars, including seven sn stars and two non-sn stars
that belong to four open clusters, for 23 different chemical species.
We used the calibration of \citet{napi93} with the Str\"omgren uvby$\beta$ photometry of
\citet{hauck-mermilliod98} to derive the fundamental parameters T$_{\rm eff}$ and log g.
These parameters were corrected for the He-weak star HD 5737 and for the HgMn star
HD 23950, based on the abnormal colour present in CP stars.
We improved them by requiring ionization and excitation equilibrium of Fe
lines. We derived the abundances by fitting synthetic and
observed spectra using the program SYNTHE, together with ATLAS9 model atmospheres.
The complete line-by-line abundances, including plots of the synthetic spectra,
are available on the web\footnote{http://icate-conicet.gob.ar/saffe/sn/Html/Salida7.html}.

We derived the rotational velocities homogeneously for our sample stars, and we agreed with
the conclusion of \citet{mermilliod83} that the sn stars present relatively low projected
rotational velocities. Although shell-like stars usually present higher rotational
velocities, the shell-like nature cannot be completely ruled out based solely on the
rotational properties.
We also compared five stars that belong to the same cluster (NGC 6475), and show that 
the sn characteristics appear in three stars with lower rotational velocity.
This apparent preference for stars with lower vsini values should be taken with caution
owing to how few objects are studied here.

We derived the abundances for the stars in our sample and compared them with the abundance values
of CP stars. We verified that approximately $\sim$40$\%$ of the sn stars studied are CP stars,
within a temperature range of 10300 K - 14500 K;
however, no clear common spectral feature or abundance value is apparent for the sn stars.
As a group they present a rather inhomogeneous chemical composition.
We also showed that not all sn stars display classical CP abundances; i.e.,
some sn stars present a $\sim$solar abundance pattern.
There is lack of a clear relationship between sn stars and CP stars.
Although the sn characteristics can coexist simultaneously with the abnormal abundances, the
chemical peculiarity does not seem to be a requirement for the appearance of the sn phenomena.
We find no clear relation between the fundamental parameters (T$_{\rm eff}$, log g, and age)
and the chemical composition of the sn stars. The small number of stars analysed could
prevent us to detect some clear trend.

We studied the possible contribution of three different processes to the broad He I lines present
in the sn stars. We used a model atmosphere trying to account for both sharp and broad lines
observed in the sn stars. Although NLTE effects could not be completely ruled out in the He I lines,
it seems that they are not directly related to the broad He I profiles observed in the sn stars.
The broad line He I 4026 {\AA} is the clearest example of the sn characteristics in our sample.
By using the appropriate Stark broadening tables, we succesfully fit this line
in four out of seven sn stars, while small differences appear in the other three stars.
We conclude that the broad He I lines that characterize the sn class could be
modelled by the usual radiative transfer process with Stark broadening, at least in some of the sn stars.
In other words, the observed broadening seems to be related to the "normal" He line formation
that originated in the atmospheres of these stars.
Studying the plots of abundance vs depth for the He I lines showed that some sn stars
are also probably stratified in He. However, a
further study of variability in the He I lines would help to determine if a possible
non-uniform He distribution could also play a role in these sn stars.


\begin{acknowledgements}
The authors thank Drs. R. Kurucz, F. Castelli, P. Bonifacio and L. Sbordone
for making their codes available to them.
We appreciate support from CONICET through grant PIP 1113-2009.
We also thank the anonymous referee whose suggestions improved the paper.

\end{acknowledgements}

\end{document}